# Galaxy formation in the *Planck* cosmology - I. Matching the observed evolution of star formation rates, colours and stellar masses

Bruno M. B. Henriques[1]*, Simon D. M. White[1], Peter A. Thomas[2], Raul Angulo[3], Qi Guo[4], Gerard Lemson[1], Volker Springel[5,6], Roderik Overzier[7]

[1] *Max-Planck-Institut für Astrophysik, Karl-Schwarzschild-Str. 1, D-85741 Garching b. München, Germany*
[2] *Astronomy Centre, University of Sussex, Falmer, Brighton BN1 9QH, UK*
[3] *Centro de Estudios de Física del Cosmos de Aragón, Plaza San Juan 1, Planta-2, 44001, Teruel, Spain*
[4] *Partner Group of the Max-Planck-Institut für Astrophysik, National Astronomical Observatories, Chinese Academy of Sciences, Beijing, 100012, China*
[5] *Heidelberger Institut für Theoretische Studien, Schloss-Wolfsbrunnenweg 35, D-69118 Heidelberg, Germany*
[6] *Zentrum für Astronomie der Universität Heidelberg, ARI, Mönchhofstr. 12-14, D-69120 Heidelberg, Germany*
[7] *Observatório Nacional/MCTI, Rua José Cristino, 77. CEP 20921-400, São Cristóvão, Rio de Janeiro-RJ, Brazil*

**ABSTRACT**
We have updated the Munich galaxy formation model to the *Planck* first-year cosmology, while modifying the treatment of baryonic processes to reproduce recent data on the abundance and passive fractions of galaxies from $z = 3$ down to $z = 0$. Matching these more extensive and more precise observational results requires us to delay the reincorporation of wind ejecta, to lower the surface density threshold for turning cold gas into stars, to eliminate ram-pressure stripping in haloes less massive than $\sim 10^{14}\,\mathrm{M}_\odot$, and to modify our model for radio mode feedback. These changes cure the most obvious failings of our previous models, namely the overly early formation of low-mass galaxies and the overly large fraction of them that are passive at late times. The new model is calibrated to reproduce the observed evolution both of the stellar mass function and of the distribution of star formation rate at each stellar mass. Massive galaxies ($\log M_*/\mathrm{M}_\odot \geqslant 11.0$) assemble most of their mass before $z = 1$ and are predominantly old and passive at $z = 0$, while lower mass galaxies assemble later and, for $\log M_*/\mathrm{M}_\odot \leqslant 9.5$, are still predominantly blue and star forming at $z = 0$. This phenomenological but physically based model allows the observations to be interpreted in terms of the efficiency of the various processes that control the formation and evolution of galaxies as a function of their stellar mass, gas content, environment and time.

## 1 INTRODUCTION

Galaxy formation theory has developed dramatically over the last three decades. Λ cold dark matter (ΛCDM) has been established as the standard model for cosmological structure formation, and its parameters have been increasingly tightly constrained by observations. In parallel, simulations of galaxy formation within this standard model have grown in complexity in order to treat more accurately the many baryonic processes that impact the evolution of the galaxy population. Semi-analytic modelling is a particular simulation method which is optimized to connect the observed properties of the galaxy population – abundances, scaling relations, clustering and their evolution with redshift – to the astrophysical processes that drive the formation and evolution of individual galaxies (e.g. White 1989; Cole 1991; Lacey & Silk 1991; White & Frenk 1991; Kauffmann et al. 1993; Cole et al. 1994; Somerville & Primack 1999; Kauffmann et al. 1999; Springel et al. 2001; Hatton et al. 2003; Springel et al. 2005; Kang et al. 2005; Lu et al. 2011; Benson 2012). Simple phenomenological descriptions of the relevant processes are needed, each typically involving uncertain efficiency and scaling parameters. These must be determined by comparison with observation or with more detailed simulations. As the range and quality of observational data have increased, so has the number of processes that must be included to model them adequately, and hence the number of adjustable parameters. In recent years, robust statistical methods have been introduced in order to sample the resulting high-dimensional parameter spaces and to determine the regions that are consistent with specific observational data sets. This development began with the work of Kampakoglou et al. (2008) and Henriques et al. (2009) and has since been extended to a wide range of models and sampling methods (Benson & Bower 2010; Bower et al. 2010; Henriques & Thomas 2010; Lu et al. 2011, 2012; Henriques et al. 2013; Mutch et al. 2013; Benson 2014; Ruiz et al. 2015).

---

* E-mail:bhenriques@mpa-garching.mpg.de

Semi-analytic modelling is designed to interpret the statistical properties of the galaxy population, and particular emphasis has always been placed on local galaxies for which abundances, scaling relations and clustering are best determined. The sharp high-mass cut-off in the observed stellar mass function can be explained by invoking efficient feedback from central black holes (Benson et al. 2003; Granato et al. 2004; Bower et al. 2006; Croton et al. 2006) but the properties of low-mass galaxies, where star formation is limited by strong stellar feedback, remain poorly reproduced by even the most recent models. Typically, the fraction of low-mass galaxies ($8.0 \leqslant \log M_*/M_\odot \leqslant 9.5$) that are no longer star forming is substantially overestimated at low redshift, as is their high-redshift abundance (Weinmann et al. 2006; Henriques et al. 2008; Fontanot et al. 2009; Guo et al. 2011, 2013; Henriques et al. 2012; Weinmann et al. 2012; Hirschmann et al. 2014). These difficulties have become more prominent in recent years as observational surveys at high redshift have improved. Massive galaxies ($\log M_*/M_\odot \geqslant 11.0$) seem to have assembled most of their mass by $z = 1$, while the abundance of low-mass galaxies grows substantially at later times (Fontana et al. 2006; Faber et al. 2007; Pozzetti et al. 2010; Marchesini et al. 2009; Ilbert et al. 2010; Marchesini et al. 2010; Ilbert et al. 2013; Muzzin et al. 2013). Furthermore, vigorous star formation is almost ubiquitous in nearby low-mass galaxies, while most massive galaxies are currently red and appear to have formed few stars since $z \sim 1$ (e.g. Kauffmann et al. 2003; Baldry et al. 2004; Brinchmann et al. 2004; Thomas et al. 2005; Arnouts et al. 2007; Drory et al. 2009; Peng et al. 2010).

In a recent paper (Henriques et al. 2013), we showed that delaying the reincorporation of gas ejected by supernova-driven winds can shift the mass assembly of dwarf galaxies to lower redshift without significantly affecting massive systems. Here, we include this change and modify additional aspects of the Munich model which affect star formation in low-mass galaxies. With respect to Guo et al. (2011) we decrease the cold gas density threshold for star formation and we eliminate ram-pressure stripping in all but the most massive dark haloes, a more drastic version of the modification advocated for similar reasons by Font et al. (2008). These changes increase the fraction of low-mass satellite galaxies that are blue at low redshift. The fraction of low-mass centrals that are blue is also increased substantially by the delayed reincorporation of wind ejecta, so together these changes produce a dwarf population which is predominantly blue at $z = 0$. We also change our model for AGN feedback to make it more efficient at late times. This is needed to ensure that galaxies around the knee of the stellar mass function are predominantly quenched by $z = 0$ despite growing significantly in number at $z \leqslant 2$. Finally, we adjust the underlying cosmological parameters of our model to correspond to the first-year *Planck* results.

By fully sampling the predicted properties for a high-dimensional space of model parameters, we demonstrate that the observed evolution of the abundance of galaxies can be reproduced as a function of stellar mass *and* specific star formation rate (sSSFR) by plausible representations of the relevant astrophysics within the standard $\Lambda$CDM structure formation model. To achieve an adequate representation of current observations it is necessary to change the modelling of a number of astrophysical processes - as we will show, our earlier models cannot fit these new data for *any* values of their parameters. Furthermore, these new data provide moderately tight constraints on all the parameters of the updated model leaving no major degeneracies. The fact that the updated model provides a good fit to the observed stellar mass functions and passive fractions over the full redshift range, $0 \leqslant z \leqslant 3$, is a significant success. It also predicts many other properties of galaxies (e.g. sizes, morphologies, gas fractions, clustering) which were not explicitly used in setting its parameters and which can be used to test it and to refine it further. Throughout this paper we will refer to all galaxy properties computed by the model, that are not directly used as inputs, as model predictions. These include the properties used to define the best-fitting set of parameters.

In a companion paper (Henriques et al. 2015b, in prep., hereafter Paper II) we analyse the processes that cause galaxies to migrate between the active and passive populations, comparing with observationally derived quenching efficiencies as a function of stellar mass, halo mass, local density and clustering, and focusing in particular on the data presented by Peng et al. (2010), Wetzel et al. (2012) and Kauffmann et al. (2013). In future work we will further test our model by comparing additional properties to available observations.

Detailed predictions for this new model are made publicly available with this paper[1]. These include snapshot and light-cone outputs with pre-computed magnitudes for various stellar population models and a wide range of broad-band filters, as well as simulated optical spectra and full star formation histories (see section 3.3 below). In the following section we briefly describe modifications in the current model with respect to previous versions. The extensive Supplementary Material presents a full description of the treatment of astrophysical processes. Section 3 describes how we set parameters in our new model. Section 4 compares its predictions to our calibrating observational data sets, while section 5 compares to additional data. We conclude with a short summary of our main results.

---

[1] The Munich models of galaxy formation can be found at http://www.mpa-garching.mpg.de/millennium.

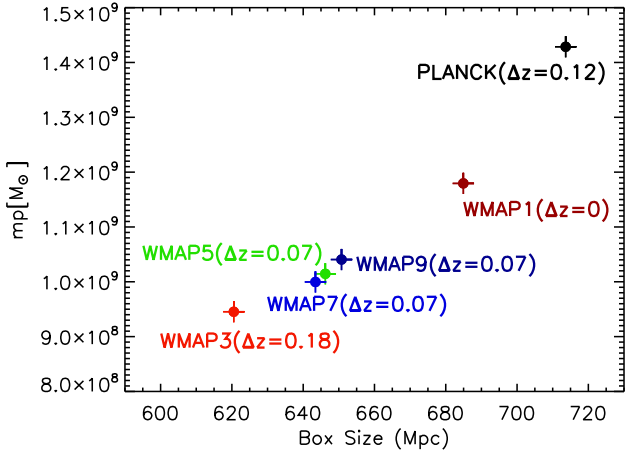

**Figure 1.** The box size and particle mass required for the Millennium Simulation to give a good representation of structure formation in different underlying cosmologies. Numbers in brackets give the difference in redshift between $z = 0$ in the target cosmology and in the *WMAP1* cosmology of the original simulation. Hence, to represent the *Planck* cosmology for this paper, the box size and the particle mass have to be scaled up by about 4 and 21%, respectively. The $z = 0.12$ output of the original simulation becomes the new $z = 0$.

# 2 UPDATES TO GALAXY FORMATION MODELLING

In this section we describe how the treatment of astrophysical processes within the Munich galaxy formation model has changed since the last publicly released catalogues which were based on Guo et al. (2011). A complete description of the current model, including more details on the newly implemented procedures, can be found in the Supplementary Material which is published along with this paper on the arXiv and is available online[2]. The changes were driven by known problems with earlier versions of this (and many other) model. Specifically, the overly early ($z \geqslant 2$) build-up of low-mass galaxies ($8.0 \leqslant \log M_*/\mathrm{M}_\odot \leqslant 9.5$) is eliminated by changing the time-scales for gas to be reincorporated after ejection in a wind; the related problems that low-mass galaxies end up too red, too old and too clustered at $z = 0$ are eliminated by additionally lowering the cold gas density threshold for star formation and removing ram-pressure effects on satellites in low-mass groups ($\log M_{200}/\mathrm{M}_\odot < 14.0$); the problem that too many massive galaxies ($\log M_*/\mathrm{M}_\odot \sim 11.0$) continue to form stars at low redshift is eliminated by assuming radio mode feedback to scale with global, rather than local, halo properties. In addition, we rescale the Millennium simulations to represent the cosmology preferred by the first-year results from *Planck*.

---

[2] A complete description of the model presented in this paper, as well as all the theoretical functional forms plotted, are available for download at http://galformod.mpa-garching.mpg.de/public/LGalaxies/

## 2.1 Simulations and cosmology

The Munich model of galaxy formation has been implemented on two simulations of the evolution of dark matter structure. Combined, the Millennium (Springel et al. 2005) and Millennium-II (Boylan-Kolchin et al. 2009) simulations provide a dynamic range of five orders of magnitude in stellar mass ($10^{7.0}\,\mathrm{M}_\odot < M_\star < 10^{12}\,\mathrm{M}_\odot$), resolving the smallest galaxies observed at $z = 0$ while also sampling the largest clusters. Over the stellar mass range $10^{9.5}\,\mathrm{M}_\odot < M_\star < 10^{11}\,\mathrm{M}_\odot$, Guo et al. (2011) found good numerical convergence between the two simulations, both for the abundance of galaxies and for their mass- and colour-dependent clustering. A similar level of convergence is found for most of the properties discussed in this paper, and throughout we will use the Millennium Simulation to derive properties for galaxies more massive than $10^{9.5}\,\mathrm{M}_\odot$ and the Millennium-II to derive properties at lower mass.

We use the technique of Angulo & White (2010), as updated by Angulo & Hilbert (2015), to scale the evolution of dark matter structure predicted by the Millennium and Millennium-II Simulations to the recently published *Planck* cosmology. Specifically, the cosmological parameters we adopt from Planck Collaboration (2014) are: $\sigma_8 = 0.829$, $H_0 = 67.3$ km s$^{-1}$Mpc$^{-1}$, $\Omega_\Lambda = 0.685$, $\Omega_\mathrm{m} = 0.315$, $\Omega_\mathrm{b} = 0.0487$ ($f_\mathrm{b} = 0.155$) and $n = 0.96$. As can be seen in Fig. 1, structural growth for this new set of cosmological parameters is as close to growth in the original Millennium (*WMAP1*) cosmology as to that in most of the cosmologies implied by subsequent *WMAP* releases. For example, in terms of structure formation, the *WMAP5* cosmology chosen for the Bolshoi (Klypin et al. 2011) and Multi-Dark (Prada et al. 2012) simulations differs from the currently preferred *Planck* cosmology by twice as much as the original *WMAP1* cosmology of the Millennium Simulations in terms of the scaling required ($\sim 9\%$ in box size and $\sim 29\%$ in particle mass compared to $\sim 4\%$ in box size and $\sim 17\%$ in particle mass), but only by half as much in terms of the time offset ($\Delta z = 0.07$ rather than $\Delta z = 0.12$). However, as shown by Wang et al. (2008), Guo et al. (2013) and Fontanot et al. (2012), the differences in cosmological parameters between all these modern determinations have a much smaller effect on galaxy properties than the uncertainties in galaxy formation physics. As a result, the change in cosmology has little impact in our conclusions.

## 2.2 Astrophysical modelling

### 2.2.1 Reincorporating wind ejecta

A number of recent studies have found that both semi-analytic and hydrodynamic simulations of the evolution of the galaxy population tend to form low-mass galaxies, $8.0 \leqslant \log M_*/\mathrm{M}_\odot \leqslant 9.5$, too early ($z \geqslant 2$), with the result that they are substantially more abundant than observed at early times (Fontanot et al. 2009; Weinmann et al. 2012; Lu et al. 2014). In Henriques et al. (2013) we showed that a possible solution to this problem is to couple strong feedback with a significantly increased delay in the time for gas to be reincorporated after ejection in a galactic wind. This substantially reduces the growth of low-mass systems at high redshift, which is then compensated by enhanced growth between $z = 2$ and 0 as the ejected gas finally returns and fuels star formation. The result is a much better match to the observed evolution of the stellar mass function.

In the current work we adopt the specific implementation proposed by Henriques et al. (2013), where the time-scale for material to be reincorporated scales directly with halo mass and is independent of redshift. A very similar dependence on parameters was found by Oppenheimer & Davé (2008) and Oppenheimer et al. (2010) in their cosmological hydrodynamics simulations, which gave reasonable fits to the evolution of the low-mass stellar mass function. In practice, this assumption means that wind ejecta from low-mass haloes take a long time to return, while ejecta from massive systems are almost immediately reincorporated. Since the return rate that we assume depends on the *current* mass of a halo rather than on its mass at the time of ejection, substantial halo mass growth can considerably accelerate the reincorporation of previously ejected material.

Firmani et al. (2010) implemented a similar dependence of gas re-accretion efficiencies in an attempt to obtain down-sizing of star formation rates in analytic models of star-forming disc galaxies in a ΛCDM universe. They argued that re-accretion alone cannot explain observed trends, which appeared to require inclusion of additional processes such as AGN feedback and a detailed model for cooling. Recently, Mitchell et al. (2014) and White et al. (2015) implemented the Henriques et al. (2013) reincorporation scheme in their own semi-analytic models. While the former show results consistent with ours, the latter found that it did not prevent the early build-up of low-mass galaxies. This presumably indicates the importance of fully exploring how this process interacts with details of the implementation of all the other physical processes that shape galaxy populations. Neistein & Weinmann (2010) and Wang et al. (2012) obtained a similar delay in the build-up of galaxies with $8.0 \leqslant \log M_*/M_\odot \leqslant 9.5$ at early times as in our new reincorporation model by choosing appropriate scalings of cooling efficiency with halo mass and redshift. We do not allow similar freedom in our own model, since we consider the cooling of diffuse gas to be one of the better understood aspects of galaxy formation, and the scalings adopted here agree reasonably well with hydrodynamic simulations (Forcada-Miro & White 1997; Benson et al. 2001; Yoshida et al. 2002).

Recently completed hydro-simulations (Vogelsberger et al. 2014; Schaye et al. 2015), have significantly improved the level of agreement between the evolution of the stellar mass function predicted by these methods and observations (Genel et al. 2014; Furlong et al. 2014; Crain et al. 2015). It would therefore be interesting to compare the gas reincorporation time-scales in these simulations with our assumptions and we plan to do so in future work.

### 2.2.2 *Star formation thresholds*

A related problem which was obvious in previous versions of the Munich model (and in many other galaxy formation models) is the fact that a large fraction of simulated galaxies with $8.0 \leqslant \log M_*/M_\odot \leqslant 9.5$ are no longer forming stars by $z = 0$ and hence are red, whereas SDSS data indicate that the great majority of real low-mass galaxies are, in fact, blue (Guo et al. 2011, 2013). The delayed build-up of these low-mass galaxies produced by our new reincorporation model reduces the discrepancy between model and observations but does not remove it completely. While low-mass central galaxies re-absorb gas at late times, thereby fuelling continued star formation, environmental effects remove the gas supply of low-mass satellites which then turn red relatively quickly.

A study of the properties of these satellites reveals that while much of their gas is removed by interactions with their host halo, they still retain a significant amount of cold gas in their discs. This gas is not transformed into stars, however, because the model adopts a gas surface density threshold for star formation. Recent observations suggest that any such threshold is lower than previously thought, and that star formation is better modelled as being linked to molecular rather than to total gas surface density, hence depending on the processes that convert HI to $H_2$ (Bigiel et al. 2008; Leroy et al. 2008). Detailed semi-analytic models for atomic to molecular gas conversion have been developed by Fu et al. (2012, 2013) and Lagos et al. (2011) and will be included in future versions of the Munich model. For this paper we just note that, in practice, stars can form from gas with surface density below our previous standard threshold, so we simply decrease this threshold by approximately a factor of 2. Satellites can then consume a larger fraction of their cold gas and so keep forming stars for longer periods.

### 2.2.3 *Environmental effects*

Although lowering the threshold for star formation decreases the number of red satellites, these remain considerably more numerous than observed. This seems to indicate that the environmental suppression of star formation is too strong in our model. A variety of processes affect satellite galaxies, but a detailed analysis indicates that the removal of hot gas reservoirs by ram-pressure stripping is already sufficient to quench star formation in satellites to below the observed levels. Indeed, suppressing this effect entirely (with no change to other processes) results in a good match to the observed fraction of passive satellites (Kang & van den Bosch 2008; Weinmann et al. 2010). Since ram-pressure stripping is *observed* to occur in rich clusters where there are substantial X-ray emitting hot gas atmospheres, we elect to retain the process in haloes with $M_{200c} > M_{R.P.} = 10^{14} M_\odot$. In lower mass groups and clusters, X-ray data show significantly lower hot gas fractions, at least in the inner regions (see Sun 2012 for a review on the subject). Given that characteristic orbital velocities are also substantially lower in such systems, it is plausible that stripping effects should be less important there, and indeed Font et al. (2008) already advocated such a reduction with respect to the findings of direct simulations of the stripping process, noting that this significantly improved the colours of dwarf galaxies in their own galaxy formation model.

The analytic equations derived by McCarthy et al. (2008) suggest that in order for satellite galaxies to retain enough hot gas to fuel continued star formation at the levels observed in low-mass groups, the hot gas surface density of the group must be 30 times smaller than that of the satellite. This requires feedback to produce large reductions in gas content and concentration in groups. More detailed work is needed to see if this can be consistent with observations and hence explain the observed lack of environmental influence on star formation. Here, we decide simply to remove the problem assuming that ram-pressure stripping does not ex-tend to low-mass groups, including the corresponding mass threshold as an additional free parameter in our Monte Carlo Markov Chain (MCMC) analysis.

In the preferred model of this paper, we thus leave all other environmental effects unchanged. In consequence, a significant fraction of satellites are quenched, even in groups where ram-pressure stripping is inactive. As in Guo et al. (2011, 2013), infall of new material stops immediately when a halo falls within $R_{200c}$ of a larger system and tidal stripping of hot gas is then turned on. The latter parallels the stripping of dark matter from the subhalo, a process which the original $N$-simulation followed explicitly. The hot gas reservoir is removed completely once a satellite loses its own halo and becomes an "orphan". The model assumes such orphans are unable to retain gas ejected by SN feedback which is moved to the hot halo of the galaxy group. Tidal forces can completely disrupt the stellar and cold gas components of orphan galaxies, which are then added to the intra-cluster light and the hot gas atmosphere of the group/cluster, respectively. Our treatment of all these processes is fully described in the Supplementary Material.

In Paper II, we will analyse the effects of our environmental physics assumptions by comparing our model in detail both with traditional autocorrelation function measurements of clustering, as in Guo et al. (2011), and with more recent studies of the dependence of galaxy properties on the mass of the halo they occupy and on their position within it, in particular, the studies by Peng et al. (2010, 2012), Wetzel et al. (2012) and Kauffmann et al. (2013). We conclude that our new model matches most but not all of the observed trends highlighted in these papers.

### 2.2.4 Radio mode feedback

The new model of Section 2.2.1 for the reincorporation of wind ejecta ensures that the abundance of galaxies below the knee of the stellar mass function can grow substantially from $z = 2$ to $0$, as observed, reflecting considerable growth in mass of the individual systems at late times. With the radio mode feedback model of Guo et al. (2011), which was taken directly from Croton et al. (2006), this late-time growth results in too many star-forming galaxies of stellar mass close to $M_\star$ and above at $z = 0$ (see Henriques et al. 2013, and Fig. 5 below). This undesirable change can be eliminated by modifying the AGN feedback model to suppress cooling and star formation more effectively at low redshift than did the original model.

We follow the methodology of Henriques et al. (2013) and include the exponents which determine the scaling of radio mode feedback with halo mass and redshift as additional parameters in a MCMC comparison with multi redshift observations of stellar mass functions and passive fractions. We find that a relatively small change to the radio mode feedback model of Croton et al. (2006) can indeed restore the crossover between predominantly star forming and predominantly passive galaxies to the observed stellar mass while retaining the substantial growth in abundance of low-mass galaxies at low redshift. Specifically, taking the heating rate to be $\dot{E} \propto M_{\rm BH} M_{\rm hot}$ rather than $\dot{E} \propto M_{\rm BH} M_{\rm hot} H(z)$ (the form used in Croton et al. 2006 and Guo et al. 2011) substantially improves the match to observation. The abundance of galaxies at $M_\star$ and below grows significantly at $z < 2$, yet star formation is quenched in most $M_\star$ objects by $z = 0$.

The analysis presented later in this paper and especially in Paper II, shows that this simple model for radio mode feedback quenches star formation in intermediate and high-mass galaxies at rates and in locations that are roughly consistent with observations. Nevertheless, even though our new implementation ensures that most massive galaxies are red and are dominated by old populations at late times, there is an indication that some may still be more actively star forming than observed.

In addition to its scaling with hot gas properties, we have changed how AGN radio mode feedback from satellite galaxies affects the hot gas of the host halo. For satellites inside $R_{200}$, any AGN energy left over after offsetting the cooling of the hot gas of the satellite is used to offset cooling of the hot gas of the host halo. This change was introduced to fix a numerical artefact identified in the dark matter merger trees. In one group of galaxies, a small object ($\log M_* / {\rm M}_\odot \sim 9.5$) centred on a low-mass subhalo ($\log M_{200} / {\rm M}_\odot \sim 12.0$) becomes the central galaxy of a massive FOF group. Lacking a massive central black hole, and suddenly increasing its assigned hot gas by orders of magnitude (from $\log M / {\rm M}_\odot \sim 11.0$ to $\log M / {\rm M}_\odot \sim 14.0$), this galaxy then experiences catastrophic growth through cooling to become the most massive galaxy in the whole simulation. This problem was already present in previous models but is accentuated in the current version, where the delayed build-up of galaxies means that their black holes grow preferentially at late time, increasing the likelihood that a low-mass satellite galaxy does not yet have a massive black hole.

Our modification ensures that, for the group where this happened, the energy released by black holes in other cluster galaxies suppresses excessive cooling of the intracluster medium. We have checked that this change only impacts the desired object.

### 2.2.5 Stellar populations and dust model

Galaxy formation models most naturally predict the evolution of physical properties of galaxies such as stellar masses, gas fractions, star formation rates, sizes, characteristic velocities and morphologies. To convert these into observables such as photometric luminosities, sizes and colours requires additional modelling of galactic stellar populations and dust content in order to calculate the emission and absorption of light at specific wavelengths, based on the age and metallicity distribution of the stars and their spatial distribution relative to interstellar dust.

Henriques et al. (2011, 2012) showed that, at least in the context of the Munich model, stellar populations that have significant emission from TP-AGB stars in intermediate age populations (e.g. Maraston 2005; Charlot & Bruzual 2007) appear required to reconcile the observed evolution of the stellar mass and $K$-band luminosity functions. For this paper, we therefore use Maraston (2005) with a Chabrier (2003) IMF as our default stellar population model. We have tested that the as yet unpublished Charlot & Bruzual (2007) model gives similar results for all properties analysed in this paper. As part of our data release we will also include lightcones with luminosities computed using the older Bruzual & Charlot (2003) model, though we caution that the rest-frame near-IR luminosities appear to be significantly underestimated in some situations in this case.

In producing photometric catalogues, we also adjusted the high redshift conversion of gas/metals into dust. In Guo et al. (2011) this fraction of gas/metals in dust was assumed to scale as $(1+z)^{-0.4}$ whereas we now adopt $(1+z)^{-1}$. This change ensures that galaxies at very high redshift have very little extinction, as observed, and results in luminosity functions for Lyman-break galaxies at $z \geqslant 5$ that match *HST* data quite well. These effects will be discussed in Paper IV of this series (Clay et al. 2015). All quantities presented in this paper are unaffected by this change. We emphasize that this is a purely phenomenological fix and may be required simply to rectify the impact of our excessively high metallicities at early times.

# 3  SETTING PARAMETERS

## 3.1  Fiducial data and MCMC procedures

The modelling framework set out above scales a high-resolution dark matter only simulation of cosmic structure formation to a specific desired cosmology (here the *Planck* cosmology) and implements a set of simplified phenomenological models that can be applied in post-processing to describe the processes that affect the formation and evolution of the galaxies themselves. Specifying a particular model requires choosing values for the parameters that determine the efficiency and the scaling of these processes. Some can be set to sufficient accuracy from physical arguments or by comparison with more detailed simulations, but the rest must be determined by fitting to some fiducial set of observational data.

There has been substantial recent progress in determining the abundance of galaxies as a function of stellar mass and star formation activity. Multiple surveys now appear to give broadly consistent results out to $z \geqslant 3$. For this paper we have therefore elected to use the observed stellar mass function of galaxies at $z = 0, 1, 2$ and $3$, together with the observed fraction of passive ("red") galaxies as a function of stellar mass at $z = 0, 0.4, 1, 2$ and $3$ to determine the parameters of our "best" model. As in previous work (Henriques et al. 2009; Henriques & Thomas 2010; Henriques et al. 2013) we combine results from all modern determinations of each quantity into a single "representative" data set with error bars representing the scatter between determinations as well as the statistical uncertainties in the individual measurements. We show these fiducial observations in Figs 2 and 5 together with the best-fitting model found by our MCMC exploration of parameter space. These plots are discussed in more detail below. The original observations underlying our fiducial data sets are shown in Appendix A.

Our MCMC procedures also parallel those of our earlier work (e.g. Henriques et al. 2013). In particular, we calculate a figure of merit, in effect an approximate likelihood, for each model given our representation of the observations by assuming that each blue point is independently sampled from a Gaussian with mean given by the model and variance given by the sum of an "observational" variance indicated by the error bar and a "theoretical" variance which we take to be negligible for the stellar mass functions and to be

$0.025^2$ for the red fractions. Note that our analysis neglects any sampling covariance between the different data points. Together with the fact that our error bars are subjective assessments of uncertainty, rather than direct estimates of sampling variance, this means that our MCMC procedure does not give statistically rigorous estimates of parameters and their uncertainties. Instead, it should be interpreted as an efficient method for delineating the region of parameter space which produces acceptable representations of the observational data.

## 3.2  Best-fitting model

For this paper we sample a significantly larger parameter space than in Henriques et al. (2013), including almost all the explicit free parameters in our semi-analytic model. An exception is the threshold between major and minor mergers. We find that there is some tension between the value of this parameter required to match observations of the fraction of red galaxies (one of our primary constraints) and that required to match galaxy morphologies. We slightly compromise the agreement with the observed red fraction of galaxies at $z = 1$ by fixing $R_{\mathrm{merge}}$ at $0.1$ in order to obtain reasonable morphologies for massive galaxies at $z = 0$. The best-fitting values for the 17 parameters included in our MCMC sampling are tabulated together with their marginalized $2\sigma$ ranges in the Supplementary Material which also contains, for reference, all the equations that define our galaxy formation model, as well as a more detailed analysis of the allowed confidence regions for its parameters and a comparison of their values with previous models. The predictions of our new best-fitting model are shown as solid red lines in Figs 2 and 5 where it can be seen to represent the fiducial observations quite well. Although our updated cosmology, together with the changes in our modelling of ejecta reincorporation, of star formation, of ram-pressure stripping and of AGN feedback induce shifts by factors of several in a number of parameters as compared, for example, to Guo et al. (2011, 2013), all parameters still remain well within their physically plausible ranges.

## 3.3  Public catalogues

With the completion of this paper we will provide public access to catalogues of data from our new best-fitting model. These will have the same format as previous data releases and will be available through the same SQL interface. As before, they will include data both for individual snapshots of the Millennium and Millennium-II simulations, and for light-cones constructed from the larger volume simulation. As part of the GALFORMOD project, we have recently expanded the functionality of these archives: a new interface allows asynchronous queries to the data base which avoid timeouts. In terms of the available predictions, the public catalogues will follow previous releases and will include the possibility to link galaxy and dark matter halo data bases, hence to analyse jointly a wide range of galaxy and dark halo properties. As previously for the Henriques et al. (2012) light-cones, we will include galaxy magnitudes for a large number of broad-band filters and for several stellar population models. A further extension of previous releases is the inclusion of full star formation histories for every galaxy (fully described in Paper III of this series, Shamshiri et al. 2015), allowing the construction of synthetic spectra, using an arbitrary dust model, in post-processing.

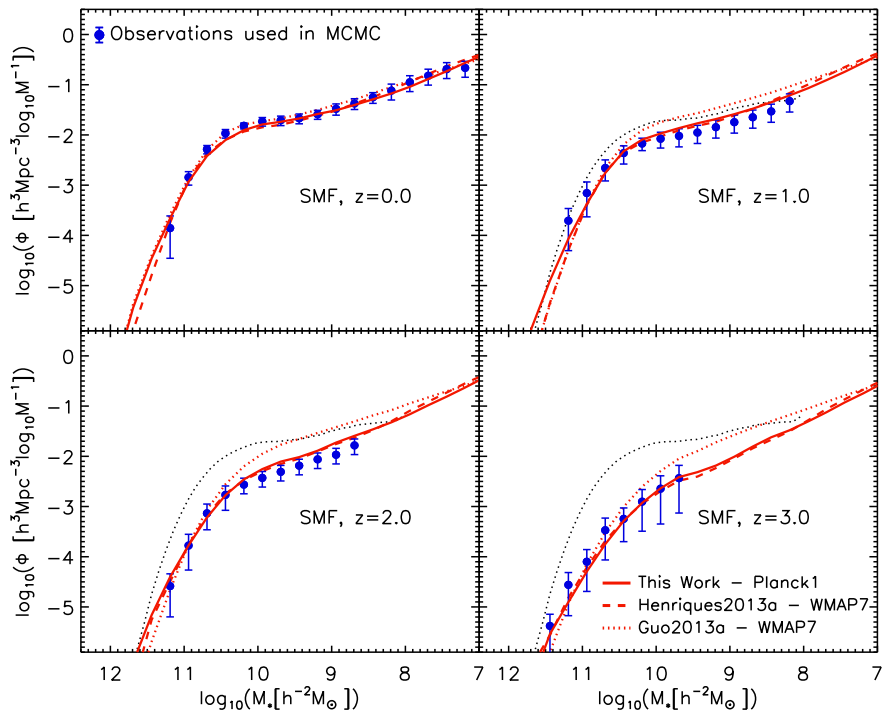

**Figure 2.** Evolution of the stellar mass function from $z = 3$ to 0. Lines show predictions from our new model (solid red), from Henriques et al. (2013) (dashed red) and from Guo et al. (2013) (dotted red). These should be compared with the blue symbols with error bars which represent the combined observational data which we use to constrain the MCMC. As described in Appendix A, the data sets we combine include SDSS (Baldry et al. 2008, Li & White 2009) and GAMA (Baldry et al. 2012) at $z \sim 0$, and Marchesini et al. (2009), Spitzer-COSMOS (Ilbert et al. 2010), NEWFIRM (Marchesini et al. 2010), COSMOS (Domínguez Sánchez et al. 2011), ULTRAVISTA (Muzzin et al. 2013, Ilbert et al. 2013) and ZFOURGE (Tomczak et al. 2014) at higher redshift. The $z = 0$ combined stellar mass function is repeated at higher redshift as a dotted black line. Here and in all subsequent plots, predicted stellar masses have been convolved with a Gaussian in $\log M_\star$, with width $0.08 \times (1+z)$, in order to account for the uncertainties in observational stellar mass determinations.

## 4 FIT TO THE FIDUCIAL OBSERVATIONS

In this section we compare our best-fitting new model in more detail with the observational data which are used as constraints in our MCMC: the evolution from $z = 3$ to 0 of the stellar mass function, and the fraction of galaxies that are "red" (i.e. no longer actively star forming) as a function of stellar mass and redshift. We pay particular attention to the fact that low-mass objects ($8.0 \leqslant \log M_*/M_\odot \leqslant 9.5$) assemble most of their mass and form most of their stars later than massive systems. This trend is often referred to as "downsizing" and seems superficially to contradict the hierarchical growth of structure expected in a $\Lambda$CDM cosmology. It has been known for some time, however, that this reversal reflects the baryonic physics of galaxy formation (e.g. De Lucia et al. 2006), and the particular implementation in this paper reproduces the observed trends in considerable detail. Dwarf galaxies form most of their stars late when material ejected in early winds is finally reincorporated. Massive galaxies ($\log M_*/M_\odot \geqslant 11.0$) are barely affected by super-nova feedback and grow quickly at high redshift, turning off when their central black holes grow big enough to suppress further accretion and so quench star formation.

In this paper we focus on the evolution of global galaxy properties, mainly stellar masses and star formation rates, distinguishing between trends at low and high stellar mass, and for passive and actively star-forming galaxies. In Paper II we will look in more detail at trends as a function of environment and their evolution with redshift. Both here and in Paper II it will turn out that the changes in the treatment of reincorporation, of star formation and of ram-pressure stripping are particularly important for dwarf galaxies, while our new treatment of AGN feedback has substantial effects only for massive objects.

In all the following sections, model stellar masses have been convolved with a Gaussian in $\log M_\star$, with width increasing with redshift, in order to mimic the uncertainties in observational stellar mass determinations. The scatter is assumed to have similar redshift dependence to that found by Ilbert et al. (2013), but larger size [a Gaussian with dispersion $0.08 \times (1+z)$ instead of $0.04 \times (1+z)$]. We note that this is slightly larger than assumed by Behroozi et al. (013a) ($0.07 + 0.04z$) and still considerably smaller than found by Conroy et al. (2009) even neglecting IMF uncertainties (0.3 at $z = 0$ and 0.6 at $z = 2$). Similar uncertainties were found by Pforr et al. (2012) and Mitchell et al. (2013) when test-ing SED fitting methods using galaxies from semi-analytic models. As discussed in Section 2.1 we use the Millennium Simulation to derive properties for galaxies more massive than $10^{9.5}$ M$_\odot$ and the Millennium-II to derive properties at lower mass.

### 4.1 Evolution of the stellar mass function - "down-sizing" in action

Fig. 2 shows galaxy stellar mass functions from $z = 3$ to 0. Results from our new model (solid red lines) are compared with results from Henriques et al. (2013) (dashed red lines) and from Guo et al. (2013) (dotted red lines), both of which assume a *WMAP7* cosmology. As explained in Appendix A, the blue symbols with error bars are the representation of the observations used in our MCMC procedures and were produced by combining a number of recent observational studies in an attempt to estimate systematic uncertainties in the constraints. Data from the various individual observational studies are plotted independently in Fig. A1. With respect to Henriques et al. (2013) we include new data sets from the UltraVISTA survey (Muzzin et al. 2013; Ilbert et al. 2013), the ZFOURGE survey (Tomczak et al. 2014) and the GAMA survey for the lowest redshift bin (Baldry et al. 2012). The new surveys are considerably deeper than previous data and have relatively large areas, allowing us to extend our fiducial data set to lower stellar masses with relatively low statistical uncertainty. When needed, we follow Domínguez Sánchez et al. (2011) and apply a correction of $\Delta M_* = -0.14$ to go from Bruzual & Charlot (2003) to Maraston (2005) stellar populations at $z \geqslant 1$.

After adjusting the parameters of our model to fit these new data, the predicted abundance of galaxies as a function of stellar mass and redshift is almost identical in our *Planck* cosmology and in the *WMAP7* cosmology of Henriques et al. (2013). As noted by Wang et al. (2008) and Guo et al. (2013), the uncertainties in galaxy formation physics produce much larger differences in galaxy masses than any change of cosmology within the currently allowed range. As already shown in Henriques et al. (2013), the longer timescales for gas reincorporation in low-mass haloes combine with stronger SN feedback to reduce the abundance of galaxies with $8.0 \leqslant \log M_*/\text{M}_\odot \leqslant 9.5$ at high redshift. The return of ejected gas at later times results in a significant build-up at $z \leqslant 1$, as required by the observational data. This late return does not drive similar low-redshift growth in massive galaxies ($\log M_*/\text{M}_\odot \geqslant 11.0$) because AGN feedback and less efficient cooling result in the production of hot gas atmospheres rather than further star formation in these systems. Overall, our new galaxy formation model (like the model of Henriques et al. 2013 before it) is able to explain the observed evolution in the stellar mass function over the last 80% of cosmic history and over the full mass range constrained by observations.

The trend for lower mass galaxies to form their stars at later times than high-mass ones (i.e. "down-sizing") can be seen more directly in Fig. 3. Here model results for the growth in mass of individual galaxies are shown for systems that have $\log_{10}(M_*[h^{-2}\,\text{M}_\odot])$ 9.75, 10.25, 10.75 and 11.25 at $z = 0$. Dashed curves give the mean stellar mass of the most massive progenitor as a function of redshift. From least massive to most massive final galaxy, the growth factors since $z = 2.5$ are 28, 18, 9 and 7, respectively, showing that giant galaxies indeed have a larger fraction of their mass in place at high redshift than dwarfs.

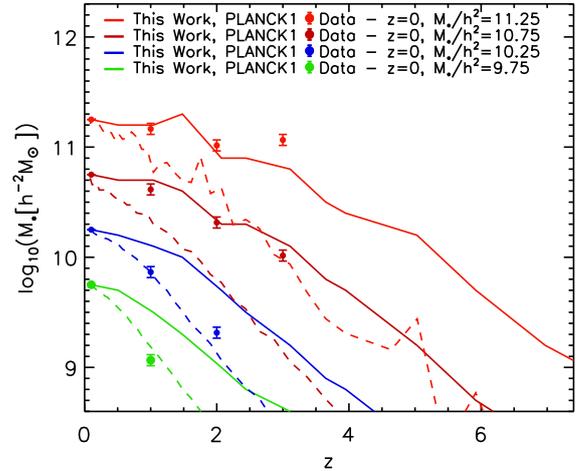

**Figure 3.** Mass growth with redshift of galaxies with different $z = 0$ mass $M_{*,0}$. The symbols with error bars are derived from the observed stellar mass functions shown in Fig. 2 by matching the cumulative abundance of galaxies $n(M_*, z)$ at each redshift $z$ to $n(M_{*,0}, 0)$. Solid lines show model predictions derived using this same abundance-matching scheme, while dashed lines show the mean stellar mass at redshift $z$ of the most massive progenitors of all galaxies that have stellar mass $M_{*,0}$ at $z = 0$.

It is not, of course, possible to observe directly the growth histories of individual galaxies, but a number of recent papers suggest that these can be inferred by assuming that the typical progenitor mass of galaxies that have stellar mass $M_{*,0}$ today, $M_*(z, M_{*,0})$, can be inferred from the "abundance-matching" equation $n(M_*, z) = n(M_{*,0}, 0)$, where $n(M_*, z)$ is the comoving abundance at $z$ of galaxies with stellar mass exceeding $M_*$ (e.g. van Dokkum et al. 2010; Brammer et al. 2011; Papovich et al. 2011). This argument neglects the scatter in assembly history among galaxies of given $M_{*,0}$ as well as the non-conservation of galaxies due to mergers. The solid curves in Fig. 3 show the result of deriving growth histories in this way in our model. It is clear that abundance matching leads to significant underestimation of the true amount of growth, giving factors of 9, 6, 4 and 3 since $z = 2.5$, two to three times smaller than the correct values. These results, for the mass growth of progenitors, are consistent with the findings of Behroozi et al. (013b) and represent a larger difference between the true mass growth and the cumulative number densities method than found for the evolution of descendants (Leja et al. 2013). Points with error bars in Fig. 3 show the result of applying this abundance-matching argument to our representation of the observational data (i.e. the blue points with error bars in Fig. 2) and hence should be compared with the solid curves. There is good agreement in the two higher mass bins, but some disagreement at lower masses.

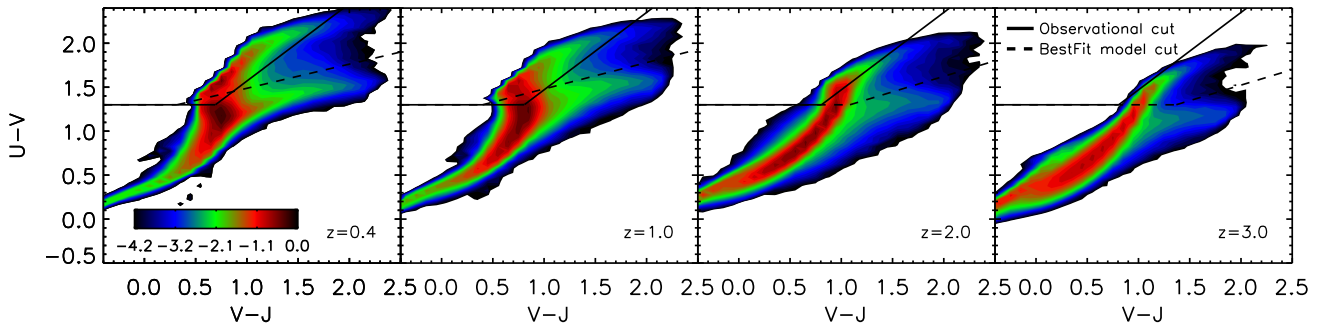

**Figure 4.** The U-V versus V-J colours of galaxies in our model. The solid line represents the observational separation into red and blue subpopulations, while the dashed line separates active and passive galaxies in our best-fitting model (see text). The colour-scale contours represent the normalized number density of galaxies in logarithmically spaced bins (from dark red high-density to dark blue low-density contours).

### 4.2 The red galaxy fraction - passive systems dominate at high mass and low redshift

The second set of constraints we use when setting the parameters of our new model are the fractions of passive galaxies as a function of stellar mass and redshift. These are obtained from observed stellar mass functions split by colour into active and passive systems, and are defined as the ratio of the estimated abundance of "red" to "red+blue" systems for stellar mass bins where estimates are available for both types of object. We start by detailing the methods used to separate red and blue galaxies in the model. Later in this section, we will compare the best-fitting model with our adopted constraints (with error bars determined by propagating the errors we assigned the original abundances) which are shown as blue points in Fig. 5. The original colour-separated mass functions from which they were derived are shown in Fig. 7. We prefer to use these passive fractions in our MCMC sampling, rather than the colour-separated mass functions themselves, in order to separate more cleanly the constraints coming from abundances from those coming from star formation activity.

A number of criteria have been proposed to separate star-forming from passive galaxies. Most involve a cut in colour or in inferred star formation rate. Recent observational studies have advocated separating galaxies using both optical and optical to near-infrared colours since this allows truly passive systems to be distinguished from dusty star-forming galaxies. Here, we use rest-frame $U - V$ and $V - J$ colours, as proposed by Muzzin et al. (2013) and also used by Ilbert et al. (2013) and Tomczak et al. (2014). Fig. 4 shows the $U - V$ versus $V - J$ rest-frame colour distribution for model galaxies at four different redshifts. A clear separation can be seen in all panels between passive galaxies (top middle), blue star-forming galaxies (bottom left) and dusty star-forming galaxies (middle right). In the observational samples, truly passive galaxies are found to lie above and to the left of the solid lines in Fig. 4. The separation is not, however, in exactly the same place in our model, presumably because of short-comings in our stellar population synthesis and dust modelling. We therefore modified the separatrix at red $V - J$ colour from the solid to the dashed line when estimating passive fractions in our model. The appropriate split appears unambiguous, except, perhaps, at $z = 3$. This division in colour correctly separates star-forming and passive galaxies in the model whereas using the original observational separation would result in passive simulated galaxies at $z \geqslant 2$ being identified as "dusty star-formers" simply because our populations synthesis/dust modelling gets their colours wrong.

At $z = 0$ we apply a cut in the magnitude-colour plane in order to match the criteria used for the observations with which we compare (Baldry et al. 2004). We again adopt a slightly different cut than in the observations when separating active and passive galaxies in the model: $u - r = 1.85 - 0.075 \times \tanh\left((M_\mathrm{r} + 18.07)/1.09\right)$. All the cuts are fixed at the separation between the two populations in the best-fitting model and unchanged during the MCMC process. In practice the cut is obtained iteratively by initially running a shorter MCMC with a colour cut based on the separation from an earlier best-fitting. For the model, we have tested that using a colour-magnitude cut, a colour-colour cut or a cut in sSFR yields very similar red and blue fractions across all redshifts. Our conclusions are thus insensitive to this choice.

Our modifications to the physics of AGN radio mode feedback, to ram-pressure stripping, and to the threshold for star formation have a substantial impact on red galaxy fractions. This can be seen in Fig. 5 which compares our representation of the observational data with predictions from our new model (the solid red lines), from the model of Henriques et al. (2013) (the dashed red lines) and from that of Guo et al. (2013) (the dotted red lines). Given that these observational data were used as constraints in our MCMC sampling, it is no surprise that the new model provides the best fit of the three. In particular, it predicts a significantly lower fraction of passive low-mass galaxies ($8.0 \leqslant \log M_*/\mathrm{M}_\odot \leqslant 9.5$) at $z \leqslant 1$ than both earlier models, and a substantially higher fraction of passive giant galaxies ($\log M_*/\mathrm{M}_\odot \geqslant 10.5$) at low redshift than the Henriques et al. (2013) model (though still somewhat fewer than observed and than in the Guo et al. (2013) model at $z = 0.4$ and 1). Passive objects are only 20% of the observed $z = 0$ population for $\log M_*/\mathrm{M}_\odot \leqslant 9.5$, and our model is now consistent with this small number.

In order to understand how our updated modelling has altered passive galaxy fractions, it is instructive to split galaxies of each type into central and satellite systems. Fig. 6 shows the stellar mass functions of active and passive galaxies at $z = 0$ as blue and red solid curves, respectively. Passive galaxies dominate the population for stellar masses above $3 \times 10^{10}$ M$_\odot$ while actively star-forming galaxies are dominant at lower masses. Dashed and dotted curves then separate each of these functions into the contributions from central and satellite galaxies, respectively.[3] Central galaxies switch from predominantly star forming to predominantly passive at $4 \times 10^{10}$ M$_\odot$ and this transition, induced by our AGN feedback prescription, is quite sharp. At $10^{11}$ M$_\odot$ more than 80% of centrals are passive, whereas at $10^{10}$ M$_\odot$ this is the case for about 20%; below $2 \times 10^9$ M$_\odot$ only about 2% of centrals are passive. Almost all passive, low-mass galaxies are satellites, but these red dwarfs are outnumbered by similar mass satellites which are still forming stars. Only for stellar masses above $5 \times 10^9$ M$_\odot$ do red satellites outnumber blue ones. This reflects the fact that most low-mass satellites are in haloes where ram-pressure stripping is no longer effective.

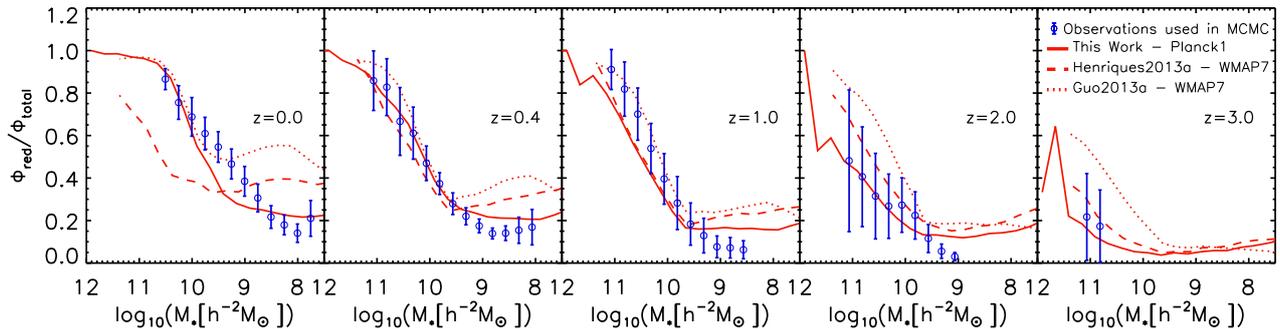

**Figure 5.** The evolution of the fraction of red (passive) galaxies as a function of stellar mass and redshift. Predictions from our new model (solid red lines), from that of Henriques et al. (2013) (dashed red lines) and from that of Guo et al. (2013) (dotted red lines) are compared with observed red fractions (blue points with error bars). These were obtained by dividing the stellar mass function of red galaxies by the sum of the red and blue stellar mass functions shown in Fig. 7. Error bars come from straightforward propagation of the errors shown in Fig. 7. A number of observational data sets were combined for this purpose: SDSS data from Bell et al. (2003) and Baldry et al. (2004) at $z = 0$ and ULTRAVISTA (Muzzin et al. 2013, Ilbert et al. 2013) and ZFOURGE (Tomczak et al. 2014) at higher redshifts.

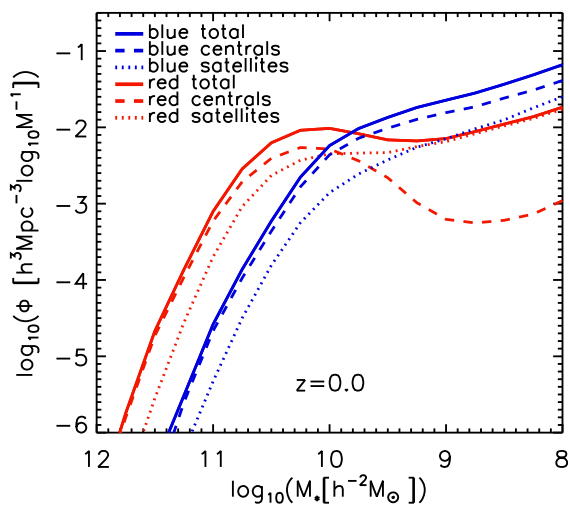

**Figure 6.** Low-redshift stellar mass functions for star-forming (blue lines) and passive (red lines) galaxies. Dashed lines give the functions for central galaxies and dotted lines the functions for satellites. The sums for each colour type are indicated by solid lines.

---
[3] Note that our definition of satellite here includes any galaxy that is not centred on the main subhalo of its friends-of-friends dark matter halo.

Most previous semi-analytic models have over-predicted the passive fraction for low-mass galaxies (e.g. Bower et al. 2006; Croton et al. 2006; Weinmann et al. 2006; Henriques et al. 2008; Henriques & Thomas 2010; Guo et al. 2011; Weinmann et al. 2012; Hirschmann et al. 2014). Our new reincorporation model ensures that central galaxies with $8.0 \leqslant \log M_*/\text{M}_\odot \leqslant 9.5$ accrete gas and continue to form stars at late times, but this alone is not sufficient to reproduce the observations (see the dashed red lines in the left-hand panels of Fig 5). The observations require a significant fraction of low-mass satellites also to remain blue. Satellites in our model do not accrete primordial gas or wind ejecta and can only remain blue if the gas they already possess at infall remains available for star formation over a long period. This forces us to suppress ram-pressure stripping and to extend star formation in the absence of new accretion, changes which are similar to those introduced by Font et al. (2008) into their own galaxy formation model for similar reasons. Note that tidal stripping of material (gas, stars and dark matter) is still present in our new model. As we will show in Paper II, our new assumptions not only reproduce the observed abundances of galaxies as a function of stellar mass, star formation rate and redshift, but also the environmental dependences and quenching time-scales inferred from observations (Peng et al. 2010, 2012; Wetzel et al. 2012; Wang et al. 2014).

The new AGN feedback implementation is needed to make sure that galaxies with $M_\star \sim 10^{10.5}$ M$_\odot$ are predominantly passive at $z < 1$ despite late-time accretion of both primordial material and wind ejecta. (Compare the dashed and solid red lines in the left-hand panel of Fig 5.) For more massive galaxies, black hole and halo masses become sufficiently large to quench star formation even earlier.

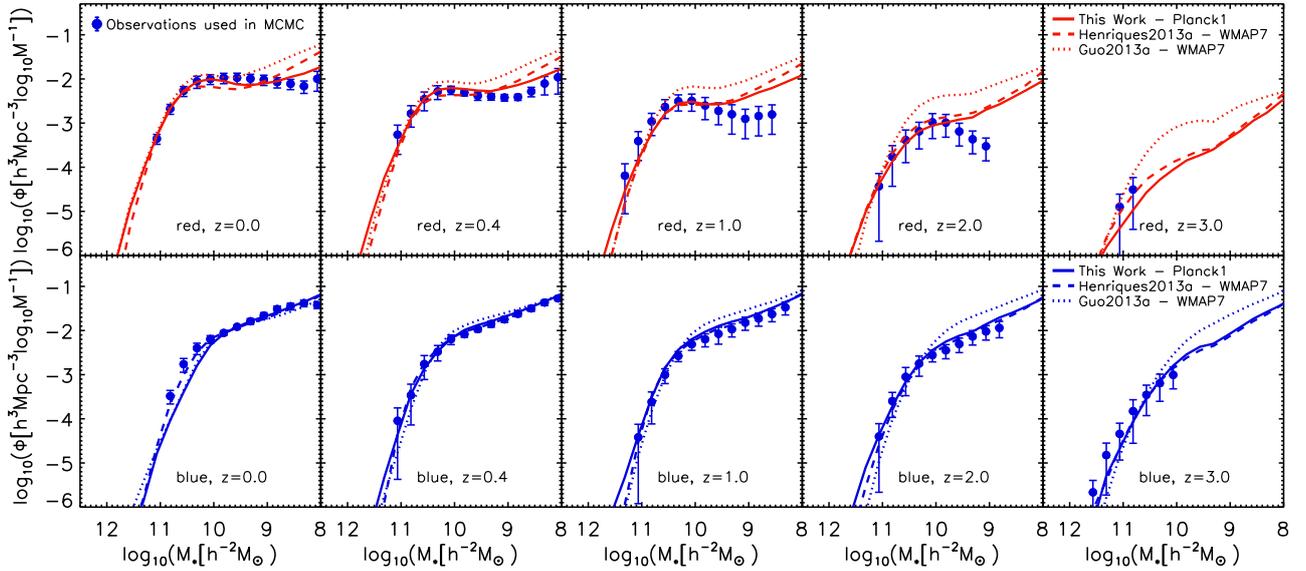

**Figure 7.** The evolution of the stellar mass functions of passive (top) and actively star-forming (bottom) galaxies from $z = 3$ to 0. Theoretical predictions from Guo et al. (2013) (the dotted lines), from Henriques et al. (2013) (the dashed lines) and from our new model (the solid lines) are compared to our representation of the available observational data (blue symbols). As discussed in Appendix A, these include SDSS data at $z = 0$ (Bell et al. 2003; Baldry et al. 2004) and ZFOURGE (Tomczak et al. 2014) and ULTRAVISTA (Muzzin et al. 2013; Ilbert et al. 2013) data at higher redshifts. The individual observations are shown in Fig. A2. As in Fig. 2, the theoretical predictions have been convolved with a Gaussian with scatter that increases with redshift as suggested by Ilbert et al. (2013) in order to represent uncertainties in the observational stellar mass estimates. Estimates originally derived using Bruzual & Charlot (2003) stellar populations have been converted to Maraston (2005) populations by applying a correction of $\Delta M_* = -0.14$ (Domínguez Sánchez et al. 2011).

# 5 FURTHER CONSEQUENCES OF DOWNSIZING

In the previous section, we focused on how and how well our new model adapts in order to fit the observational data which we use directly as constraints, namely the abundance and passive fraction of galaxies as functions of stellar mass from $z = 3$ to 0. Here, we analyse related quantities which were not used in our MCMC sampling in order to clarify the physical implications of the observed phenomenology. In particular: we study stellar mass functions separated explicitly into active and passive systems, together with the evolution of the implied cumulative mass and number densities; we compare observed and model distributions of colour, sSFR and luminosity-weighted stellar age for low-redshift galaxies; we analyse the evolution of the "star-forming main sequence" of galaxies; and we look at the evolution of the mean cosmic star formation rate density (the Lilly-Madau diagram). A detailed comparison of observations with predicted black hole masses, cold-gas fractions, metallicities, morphologies and AGN feedback rates can be found in the Supplementary Material. Comparisons of clustering predictions with observations will be presented in Paper II.

## 5.1 Abundance evolution for active and passive galaxies

In Fig. 7 we separate the stellar mass functions already plotted in Fig. 2 into passive ("red", upper panels) and actively star-forming ("blue", lower panels) systems using the colour cuts outlined in Section 4.2. The blue symbols are our representation of the results from several recent observational surveys, as detailed in Appendix A, and were used to estimate the passive fractions shown as blue symbols in Fig. 5. They are compared with model predictions from Guo et al. (2013) (dotted lines), from Henriques et al. (2013) (dashed lines) and from the model of this paper (solid lines). In this figure the differences between our new model and Henriques et al. (2013) appear significant only at low redshift and are relatively small. Both models clearly represent the observations better than Guo et al. (2013), predicting fewer low-mass galaxies at early times and fewer low-mass, passive galaxies at all times. At low redshift, our new model has fewer of these $8.0 \leqslant \log M_*/M_\odot \leqslant 9.5$ passive galaxies and more passive giants ($\log M_*/M_\odot \geqslant 11.0$) than Henriques et al. (2013) and is in reasonable agreement with the observations, although all models appear to predict more passive dwarfs than are observed at redshifts 1 and 2. The effect of our new AGN feedback is evident in the more rapid buildup of passive galaxies around the knee of the stellar mass function at late times.

Both in the observations and in our new model the stellar mass function of star-forming galaxies evolves very little between $z = 2$ and 0 and has a steep low-mass slope, whereas the mass function of passive galaxies has a much shallower low-mass slope and grows in amplitude by an order of magnitude while keeping the characteristic stellar mass at its knee almost constant. This behaviour has been noted previously (e.g. Bundy et al. 2006; Bell et al. 2007; Ilbert et al. 2010) in particular by Peng et al. (2010) who used it as the basis of a simple, toy model for galaxy formation. It implies that the growth of galaxies through star formation and mergers is being balanced by quenching processes that move galaxies from the active to the passive population. As noted by Peng et al. (2010), the constancy of the characteristic mass of passive galaxies implies that quenching typically occurs when galaxies grow to a well-defined stellar mass which is independent of time. In our physical model, this characteristic stellar mass is the minimum value for which feedback from the central supermassive black hole is able to offset cooling and accretion from the hot gas halo. As the passive galaxy stellar mass function grows in amplitude relative to that of active galaxies, the difference in shape between the two functions means that passive galaxies first start to dominate the population at high mass, and that the cross-over between active and passive domination moves steadily to lower stellar mass at later times.

A somewhat different view of this behaviour can be seen in Fig. 8. The upper panel shows comoving stellar mass density as a function of redshift for the galaxy population as a whole (black symbols and curve) and separated into the contributions from passive and active systems (red and blue symbols and curves, respectively). Symbols and shaded regions are obtained from our representation of the observed stellar mass functions and are integrated down to the observational completeness limits in each case. Predictions from our new model are shown integrated over all masses (dashed curves) and down to the same redshift- and colour-dependent completeness limits as the observation (solid curves). The stellar mass density in active systems is independent of redshift, once the variation in the completeness limit is accounted for, whereas the stellar mass density in passive systems increases by more than an order of magnitude to become dominant at $z = 0$. The lower panel of Fig. 8 shows cumulative comoving number densities of active and passive galaxies above three stellar mass thresholds as a function of redshift. Above the highest threshold, passive and active galaxies are already equally abundant at $z \sim 2$. Above the intermediate threshold, they are equally abundant at $z \sim 1$, while above the lowest threshold they only become equally abundant around $z \sim 0$. In both panels our new model is in qualitative agreement with the observational trends and appears to agree quantitatively within the considerable uncertainties.

## 5.2  Colours, sSFRs and ages of low-$z$ galaxies

At low redshift the distributions of colour, sSFR and stellar age can be measured robustly for galaxies over a very wide mass range. In Fig. 9 we show $u$-$i$ colour distributions, $r$-band luminosity-weighted age distributions and sSFR distributions for SDSS galaxies split into eight disjoint bins spanning four orders of magnitude in stellar mass, $8.0 \leqslant \log M_*/M_\odot \leqslant 12.0$. The observational data are shown as solid black histograms. sSFR estimates were derived as in Brinchmann et al. (2004), including the later corrections of

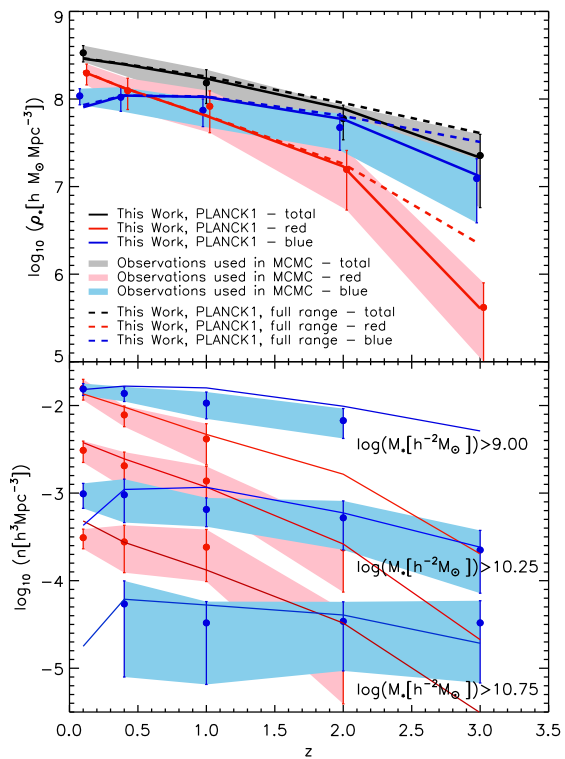

**Figure 8.** Upper panel: evolution of the comoving stellar mass density in passive (red) and active (blue) galaxies and in the full population (black). Symbols with error bars and the associated shaded regions are obtained from our compilation of observational results by integrating above the adopted completeness limit. Solid lines are predictions from the model of this paper integrated over all masses (dashed lines) and above the same completeness limits as the observations (solid lines). Lower panel: cumulative comoving number densities of active and passive galaxies above three stellar mass thresholds. Again, symbols and the associated shaded regions give observational results, while the solid lines are predictions from the model of this paper.

Salim et al. (2007). In order to match the observed distribution of unclassified galaxies with no emission lines, that were assigned a value of SFR based on SED fitting, model galaxies with $\log(sSFR) \leqslant -12$ yr$^{-1}$ have been assigned a random Gaussian value centered at $\log(sSFR) = -0.3 \log(M_*) - 8.6$ and with dispersion 0.5. Stellar masses are taken from Kauffmann et al. (2003) and stellar ages were derived as in Gallazzi et al. (2005). All observational data were downloaded from the MPA-JHU catalogue[4] and a $1/V_{max}$ factor was applied when accumulating histograms in order to produce volume-limited statistics. As in previous plots, these observational data are compared with predictions from our new model, from Henriques et al. (2013) and from Guo et al. (2013).

For masses above $10^{10}$ M$_\odot$ there are clear differences between the three models. The delayed reincorporation of wind ejecta introduced by Henriques et al. (2013) produces an increase in the number of star-forming galaxies which is visible in the three left-most lower plots as a peak around

---
[4] http://www.mpa-garching.mpg.de/SDSS/

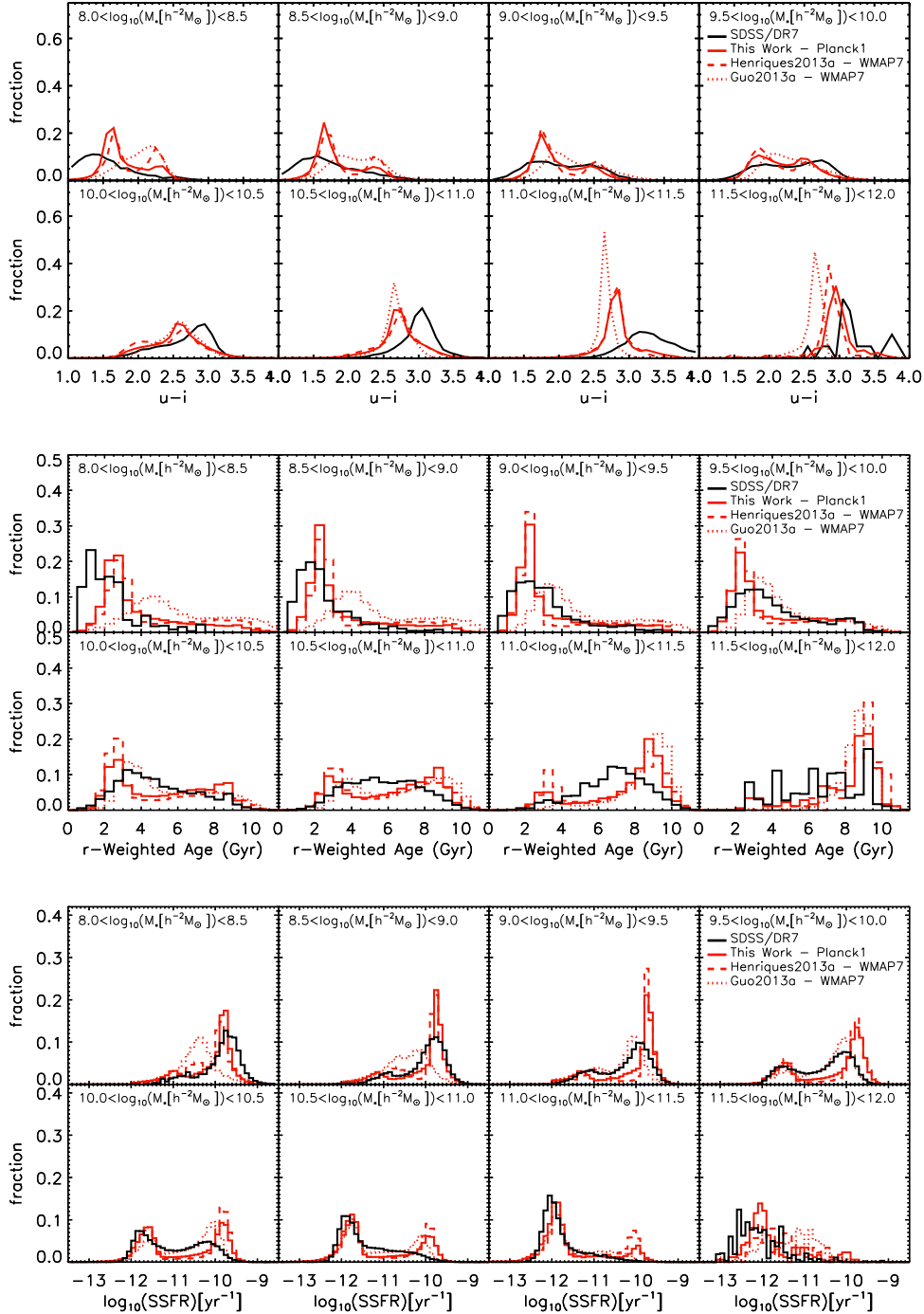

**Figure 9.** *u-i* colour (top panels), *r*-band weighted age (middle panels) and sSFR (bottom panels) distributions at $z = 0.1$ in eight disjoint stellar mass bins. Observational estimates from the SDSS are shown as black solid histograms and are compared with predictions from our new model (solid red histograms), from Henriques et al. (2013) (dashed red histograms) and from Guo et al. (2013) (dotted red histograms). sSSFR estimates for SDSS come from Brinchmann et al. (2004) and Salim et al. (2007), and luminosity-weighted stellar age estimates from Gallazzi et al. (2005).

1 to 2 Gyr in the age distributions and around $10^{-10}$yr$^{-1}$ in the sSFR distributions. These peaks are less pronounced in the earlier model of Guo et al. (2013). As a result of this shift, Henriques et al. (2013) under-predicted the fraction of passive galaxies at high mass (see Fig. 5). Our updated treatment of feedback from radio mode AGN corrects this problem (although not completely) by quenching galaxies more strongly at later times. Its effects are evident in the reduction in the fraction of objects in these peaks in the new model. At high mass, galaxy colours are significantly redder in the two more recent models than in Guo et al. (2013) because of the change in population synthesis model and the larger value adopted for the nucleosynthetic yield. Discrepancies between model and observation nevertheless remain, particularly at intermediate masses, suggesting that our treatment of the transition between the low- and high-mass regimes, where the properties of galaxies change dramatically, still needs to be improved. In this range, the model distributions are much more clearly bimodal than those of SDSS galaxies. Another important factor is the treatment of dust which might also significantly contribute to this behaviour.

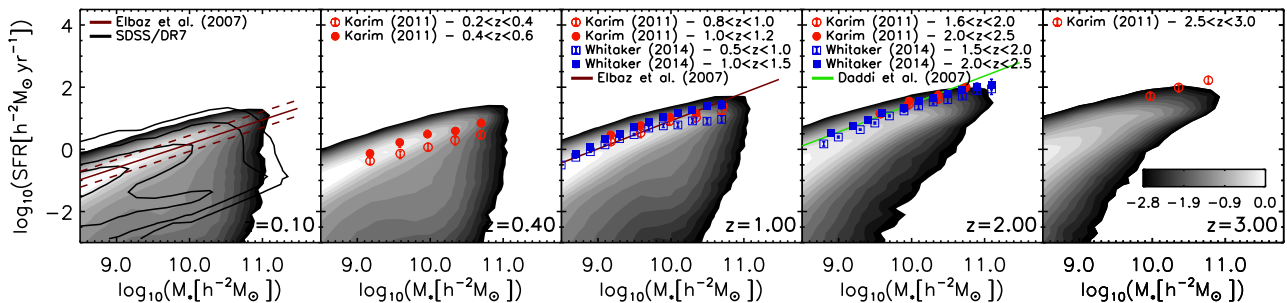

**Figure 10.** Star formation rate plotted against stellar mass from high to low redshift (right to left). The distributions predicted by the model of this paper are shown in grey-scale and are compared with symbols and straight lines representing the ridge-line of the observed "star-forming main sequence" as inferred in recent studies that are indicated individually in each panel. At $z = 0.1$ we show the distribution inferred from SDSS with solid logarithmically spaced contours. The grey-scale contours represent the normalized number density of galaxies in logarithmically spaced bins (from light grey high-density to dark grey low-density contours).

For low-mass galaxies ($8.0 \leqslant \log M_*/M_\odot \leqslant 9.5$), the improved agreement of our new model with the SDSS data is quite clear. The peaks at red colours, old ages and low sSFR which were present to varying degrees in both the Henriques et al. (2013) and Guo et al. (2013) models are now strongly suppressed. Most dwarf galaxies are indeed blue, young and star forming, as observed, with only a small fraction of red/passive galaxies remaining. This reflects both the later reincorporation of ejecta, which ensures that almost all central galaxies are blue, and our altered assumptions about ram-pressure stripping and star formation thresholds, which allow a majority of low-mass satellite galaxies to remain blue. Although some disagreements of detail remain, the new model is in reasonable qualitative agreement with the inferred star formation rates of local galaxies over the full observed stellar mass range.

### 5.3 Evolution of the star-forming main sequence and the cosmic star formation rate density

In recent years, deep surveys of the galaxy population have identified a relation between star formation rate and stellar mass which holds for the bulk of star-forming systems. This "main sequence" relation is close to a direct proportionality (e.g. Peng et al. 2010) and has relatively modest scatter at any given redshift, but its amplitude evolves strongly with time. The population of passive objects falls below this main sequence and has long been studied in detail at low redshift (e.g. Baldry et al. 2004; Kauffmann et al. 2004; Brinchmann et al. 2004). Recent increases in sensitivity have made similar analyses possible at high redshift, where these passive galaxies are the ones counted in the "red" stellar mass functions of Fig. 7 which show them to be subdominant by number, even at high mass, for $z \geqslant 2$ (see also Fig. 8)

Fig. 10 compares the redshift evolution of the relation between stellar mass and star formation rate in our new model (the grey-scale) with different observational determinations of the ridge-line of the star-forming main sequence (the solid lines and red symbols). The observed and model relations have similar slopes and evolve similarly with redshift. Both decrease in amplitude by well over an order of magnitude between $z = 2$ and 0. In addition, the scatter in the low-redshift relation is also quite similar in the model and in the data (see Fig.9). Small discrepancies between theory and observations remain. These seem consistent with the calibration uncertainties of different observational star formation rate indicators (see Kennicutt & Evans 2012 for a review on the subject) and with the detailed comparison of star formation rates in different models recently performed by Mitchell et al. (2014).

The cosmic star formation rate density is defined as the volume averaged sum of all ongoing star formation at any given time. Its integral from very early times down to redshift $z$ should be equal to the integral of the mass-weighted stellar mass function at $z$, once mass loss during stellar evolution is accounted for. In practice, however, the estimation of stellar mass and star formation rate densities from observational data is subject to large uncertainties and deviations between observational determinations of these two quantities are to be expected (Hopkins & Beacom 2006; Wilkins et al. 2008). As a result, a model that correctly predicts the observed evolution of the stellar mass function may not match observational estimates of the evolution of the star formation rate density.

In Fig. 11 we compare model predictions to observational estimates of the star formation rate density from $z = 9$ to 0. As in previous figures we show results for the new model of this paper as a solid red line, results for the model of Henriques et al. (2013) as a dashed red line, and results for the Guo et al. (2013) model as a dotted red line. The observations are taken from COSMOS (Karim et al. 2011), the Bouwens et al. (2012) sample of Lyman-break galaxies, combined Herschel and *HST H* band-selected catalogues (Schreiber et al. 2015) and the Behroozi et al. (013a) compilation. The observed rate has a broad peak at relatively

## 6 SUMMARY AND CONCLUSIONS

We have updated the cosmological parameters underlying our galaxy formation model to the values preferred by the first analysis of *Planck* CMB data, while modifying our treatment of baryonic processes to address two major problems identified in earlier modelling, namely the over-prediction of the abundance of low-mass galaxies ($8.0 \leqslant \log M_*/M_\odot \leqslant 9.5$) at redshifts $z \geqslant 1$ and the overly large passive fraction predicted among low-redshift dwarfs. We use recent observational estimates of the abundances and passive fractions of galaxies over the stellar mass range $8.0 \leqslant \log M_*/M_\odot \leqslant 12.0$ and the redshift range $0 \leqslant z \leqslant 3$ as constraints on our modelling, using MCMC procedures to identify the thresholds, scaling exponents and efficiencies needed for our treatment of baryonic processes to match the observations.

Relative to the most recent of our previous publicly-released models (Guo et al. 2011, 2013) matching these observations required us to delay the return of material ejected in galactic winds (as in Henriques et al. 2013), to weaken ram-pressure stripping in low-mass haloes with $\log M_{200}/M_\odot < 14.0$ (as advocated by Font et al. 2008, for their own galaxy formation models), to lower the gas surface density threshold for star formation, and to make radio mode feedback from AGN more efficient at late times. With these changes, our new model reproduces our fiducial observations well over their full stellar mass and redshift ranges. In particular, it matches both the observed abundance of low-mass galaxies at $z \geqslant 1$ and the observed sharp, low-redshift transition between predominantly star-forming systems at low mass, $8.0 \leqslant \log M_*/M_\odot \leqslant 9.5$, and predominantly passive galaxies at high mass, $\log M_*/M_\odot > 10.5$. For low-redshift galaxies, the detailed distributions of colour, sSFR, and luminosity-weighted stellar age are matched reasonably well across the entire stellar mass range, $8.0 \leqslant \log M_*/M_\odot \leqslant 12.0$. In addition, the evolution of the mean cosmic star formation rate density over the range $0 < z < 9$ is reasonably well reproduced, once possible calibration uncertainties are allowed for.

Our new model embeds simple but plausible representations of the physical processes known to influence galaxy formation and evolution in the structure formation framework of the concordance $\Lambda$CDM model, yet it behaves in a very similar way to the simple toy model which Peng et al. (2010) introduced to interpret the observed evolution of stellar mass functions split into star-forming and passive systems. At each redshift, there is a well-defined star-forming main sequence along which sSFR varies only weakly. The stellar mass function of star-forming galaxies has a steep low-mass slope and evolves very little with redshift, whereas that of passive galaxies has a much flatter low-mass slope and grows strongly in amplitude, but weakly in characteristic mass, with decreasing redshift. As a result, passive galaxies first come to dominate the population at high mass ($\log M_*/M_\odot \sim 11.3$ at $z \sim 2$) and the transition between active and passive domination shifts to progressively lower stellar mass at later times, dropping to $\log M_*/M_\odot \sim 10.0$ by $z = 0$. Peng et al. (2010) noted that fitting the data with their toy model required quenching of star formation to occur near a characteristic stellar mass which is independent of redshift. In our physical model, this characteristic stellar low redshift ($z \sim 2$ to $3$) and declines significantly by $z = 0$ but also to higher redshift. The prediction of these general features can be considered as one of the first significant successes of semi-analytic modelling of galaxy formation in a CDM universe (White 1989). Our new model matches the overall shape of the observed relation reasonably well although it is not peaked enough at $z = 2$. It seems that, despite fully matching the most recent observations of the stellar mass function from $z = 3$ to $0$, we predict a milder decrease in the integrated star formation rate density than observed. There is thus some tension between the observational determination of these two quantities (Whitaker et al. 2014; Leja et al. 2015). Similar discrepancies were found by Furlong et al. (2014) when looking at the Schaye et al. (2015) numerical simulations.

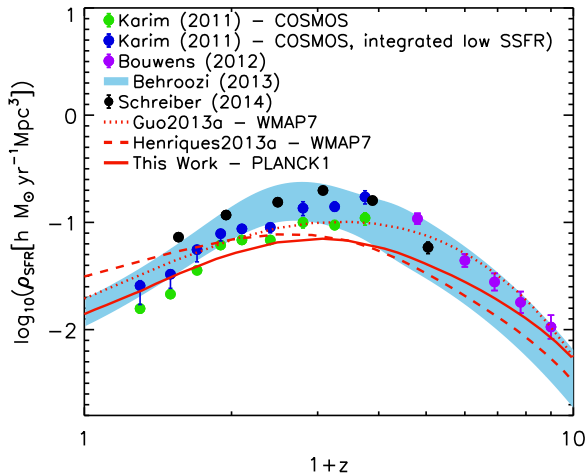

**Figure 11.** The evolution with redshift of the comoving density of cosmic star formation. The new model (solid red line), that of Henriques et al. (2013) (dashed red line) and that of Guo et al. (2013) (dotted red line) are compared with observational data from Karim et al. (2011), Bouwens et al. (2012), Schreiber et al. (2015) and Behroozi et al. (013a).

We note that our modified treatment of radio mode feedback is responsible for the substantially larger drop in the star formation rate density at $z < 2$ than in the Henriques et al. (2013) model. The change from *WMAP7* to *Planck* cosmology results in higher halo accretion rates at early times and higher star formation rate densities at $z > 2$.

As pointed out by Schaye et al. (2010) and also seen in the MCMC analysis of Henriques et al. (2013), the high-redshift star formation rate density is mostly determined by the accretion of primordial material on to haloes with virial temperatures for which cooling is efficient. Below $z = 2$, the slowing of the cosmological accretion rate combines with a lengthening of characteristic cooling times to produce a global decrease in star formation rates. In addition, at later times AGN feedback and environmental quenching mechanisms also contribute to the decrease in the integrated star formation density, moving galaxies from the main star-forming sequence into the passive population.

mass turns out to be the minimum value for which feedback from the central supermassive black hole is able to offset cooling and accretion from the hot gas halo.

In order to match the low passive fraction measured in low-redshift dwarf galaxies, we had to reduce environmental effects on dwarf satellites so that at least half of them are still star forming by the present day. This has an impact, of course, on the clustering of galaxies as a function of star formation activity, an issue which we will address in some detail in Paper II, where we will show that the updates to our modelling also substantially improve the extent to which it matches observations of such "environmental quenching". Although many aspects of our baryonic modelling remain crude, and there are still some quantitative discrepancies with observations, we believe that the new model of this paper not only updates that of Guo et al. (2011) to the currently preferred *Planck* cosmology, but also cures the principal discrepancies with observations which were discovered in the earlier work without significantly damaging any of its successes. We therefore anticipate that publicly released data catalogues for the new model will be of considerable use for interpreting a broad variety of observations of the evolution and clustering of the galaxy population.

**ACKNOWLEDGEMENTS**

This work used the DiRAC Data Centric system at Durham University, operated by the Institute for Computational Cosmology on behalf of the STFC DiRAC HPC Facility (www.dirac.ac.uk). This equipment was funded by BIS National E-infrastructure capital grant ST/K00042X/1, STFC capital grant ST/H008519/1, and STFC DiRAC Operations grant ST/K003267/1 and Durham University. DiRAC is part of the National E-Infrastructure. The work of BH, SW and GL was supported by Advanced Grant 246797 "GAL-FORMOD" from the European Research Council. PT was supported by the Science and Technology Facilities Council [grant number ST/I000652/1]. GQ acknowledges support from the National basic research programme of China (973 programme under grant No. 2009CB24901), the Young Researcher Grant of National Astronomical Observatories, CAS, the NSFC grants programme (No. 11143005), as well as the Partner Group programme of the Max Planck Society. VS acknowledges support by the Deutsche Forschungsgemeinschaft through Transregio 33, "The Dark Universe". The authors thank Ivan Baldry, Olivier Ilbert, Alexander Karim, Cheng Li, Danilo Marchesini and Adam Muzzin for providing their observational data, Eric Bell, Jarle Brinchmann, Scott Clay, Guinevere Kauffmann, Bryan Terrazas, Loic Le Tiran, Jun Toshikawa, Stephen Wilkins and Rob Yates for useful discussions and Claudia Maraston, Gustavo Bruzual and Stephan Charlot for providing their stellar populations synthesis models.

# APPENDIX A: COMPARING MODELS AND OBSERVATIONS

In this Appendix we define the figure of merit used to assess the level of agreement between our models and the observational data with which we constrain them. In addition, we show the individual observational data sets and describe how they are combined to give constraints with realistic uncertainty estimates which are suitable for MCMC exploration of the high-dimensional parameter space of our models. A more detailed description of our methods can be found in Appendix 3 of Henriques et al. (2013).

## A1  Figure of merit

Our figure of merit for each model is its "likelihood", given the constraining observations. This is computed assuming individual data points to be independently and normally distributed around the model prediction with variance corresponding to the sum in quadrature of observational and theoretical uncertainties estimated as detailed below. The observational uncertainties are dominated by systematics rather than by sampling noise, so neither the Gaussian assumption nor the precise variance can be rigorously justified. In addition, both types of uncertainty are expected to be strongly correlated between data points. As a result, our figure of merit is not a true likelihood and our MCMC analysis should be interpreted as indicating acceptable regions of parameter space, rather than as defining posterior probability distributions in a rigorous Bayesian sense (see Benson 2014, for related discussion).

Our procedure is similar to that outlined in Henriques et al. (2013). We again use observed stellar mass functions at a series of redshifts ($z = 0, 1, 2$ and $3$) to constrain galaxy abundances, but, rather than using $K-$ and $B-$band luminosity functions to constrain the relative numbers of passive and actively star-forming galaxies, we here use direct estimates of the passive fraction as a function of stellar mass at $z = 0, 0.4, 1, 2$ and $3$. These are obtained from stellar mass functions for galaxy samples split into active and passive subsets according to colour-colour plots like Fig. 4. This change makes our analysis less sensitive to the details of stellar population synthesis models and separates constraints on galaxy abundance more clearly from constraints on galaxy activity.

Given the above assumptions our figure of merit for each model (its "likelihood") can be computed as:

$$\mathcal{L} \propto \exp\left(-\chi^2/2\right), \quad (A1)$$

where $\chi^2$ is given by:

$$\chi^2 = \sum_{i,j} \frac{(\phi_{i,j} - \tilde{\phi}_{i,j})^2}{\Delta\phi_{i,j}^2 + \Delta\tilde{\phi}_{i,j}^2} + \sum_{i,j} \frac{(f_{i,j} - \tilde{f}_{i,j})^2}{\Delta f_{i,j}^2 + \Delta\tilde{f}_{i,j}^2}. \quad (A2)$$

The first sum on the r.h.s. of this equation involves the stellar mass functions $\phi$ (here defined as the logarithm of the number density of galaxies of the relevant mass and redshift), while the second involves the passive fractions $f$. In each case, the index $i$ enumerates the redshifts used, while $j$ enumerates the stellar mass bins at each redshift. Observational quantities and their adopted uncertainties are indicated by a tilde, while model predictions and their uncertainties have no tilde. The uncertainty in a quantity $x$ is denoted by $\Delta x$.

We compute passive fractions as a function of stellar mass for all modern surveys for which the authors have explicitly estimated stellar mass functions separately for active and passive systems. As discussed in the next subsection, we combine these independent estimates to obtain the stellar mass functions, $\tilde{\phi}_B$ and $\tilde{\phi}_R$ and associated uncertainties shown in Fig. 7. These functions are then combined to give the passive fractions we use as constraints

$$\tilde{f}_{i,j} = \frac{\tilde{\phi}_{R;i,j}}{\tilde{\phi}_{R;i,j} + \tilde{\phi}_{B;i,j}}. \quad (A3)$$

The uncertainties in the passive fractions are straightforwardly obtained from those in the stellar mass functions by standard error propagation,

$$\Delta\tilde{f} = \frac{\tilde{\phi}_R \tilde{\phi}_B}{(\tilde{\phi}_R + \tilde{\phi}_B)^2} \sqrt{(\Delta\tilde{\phi}_R/\tilde{\phi}_R)^2 + (\Delta\tilde{\phi}_B/\tilde{\phi}_B)^2}, \quad (A4)$$

where the index pair $i, j$ has been suppressed on all quantities for clarity.

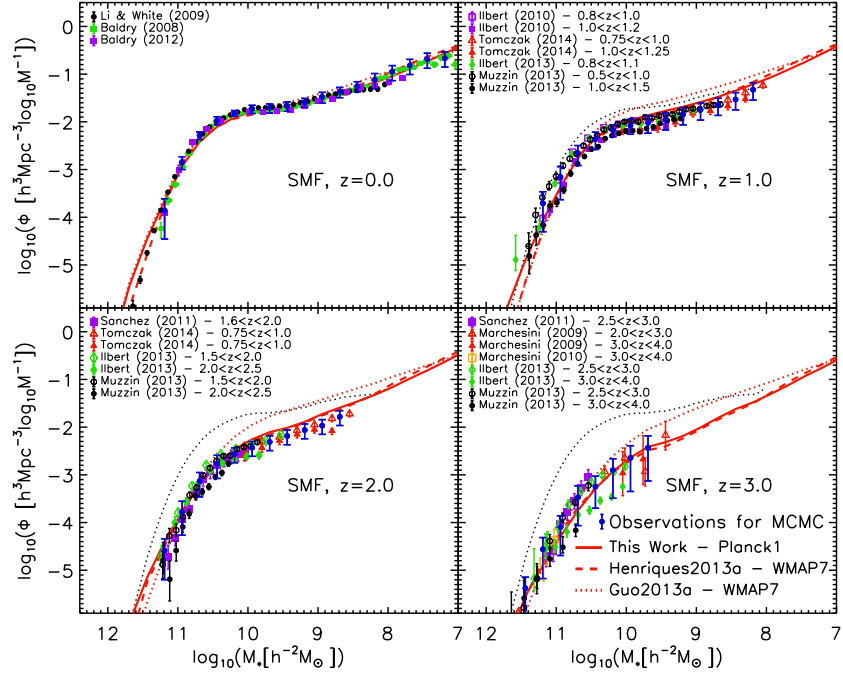

**Figure A1.** Evolution of the stellar mass function from $z = 3$ to 0 as in Fig. 2, except with the data points for the individual underlying surveys also shown. These surveys include SDSS (Baldry et al. 2008, Li & White 2009) and GAMA (Baldry et al. 2012) at $z \sim 0$ and Marchesini et al. (2009), Spitzer-COSMOS (Ilbert et al. 2010), NEWFIRM (Marchesini et al. 2010), COSMOS (Domínguez Sánchez et al. 2011), ULTRAVISTA (Muzzin et al. 2013, Ilbert et al. 2013) and ZFOURGE (Tomczak et al. 2014) at higher redshifts. All mass estimates at $z > 0$, except Domínguez Sánchez et al. (2011) and Muzzin et al. (2013) have been shifted by -0.14 dex to convert from Bruzual & Charlot (2003) to Maraston (2005) stellar populations (Domínguez Sánchez et al. 2011). The $z = 0$ results of Li & White (2009) are repeated at all redshifts as a black dotted line.

Finally we assume the theoretical uncertainty on the predicted passive fractions to be $\Delta f_{i,j} = 0.025$, based on the scatter in the passive fraction among the tree subsamples used in our MCMC analysis, and we neglect the theoretical uncertainty in the stellar mass functions, setting $\Delta \phi_{i,j} = 0$.

### A2 Individual observational data sets

As in Henriques et al. (2009), Henriques & Thomas (2010) and Henriques et al. (2013) we combine multiple determinations of each stellar mass function, using the scatter among them to indicate the systematic uncertainties which appear in most cases to be larger than those purely due to count statistics. For each redshift range and for each stellar mass bin we take a straight average of the different data sets and assume the $1\sigma$ uncertainty to be half of the maximum to minimum range. By not weighting the averages we attempt to recognise the fact that systematic errors can affect large and small surveys in similar ways. However, we emphasize that this estimate of uncertainties is crude, and that in the presence of systematics any comparison between theory and observations is essentially qualitative. Formal levels of agreement should thus be treated with considerable caution.

Our adopted constraints are shown together with the individual data sets on which they are based in Figure A1 for the overall stellar mass function and in Figure A2 for the stellar mass functions of passive and actively star-forming galaxies. The constraints we use in our MCMC sampling are shown as blue dots with error bars while other data points represent published observational estimates from the individual surveys. Theoretical predictions are shown as red lines.

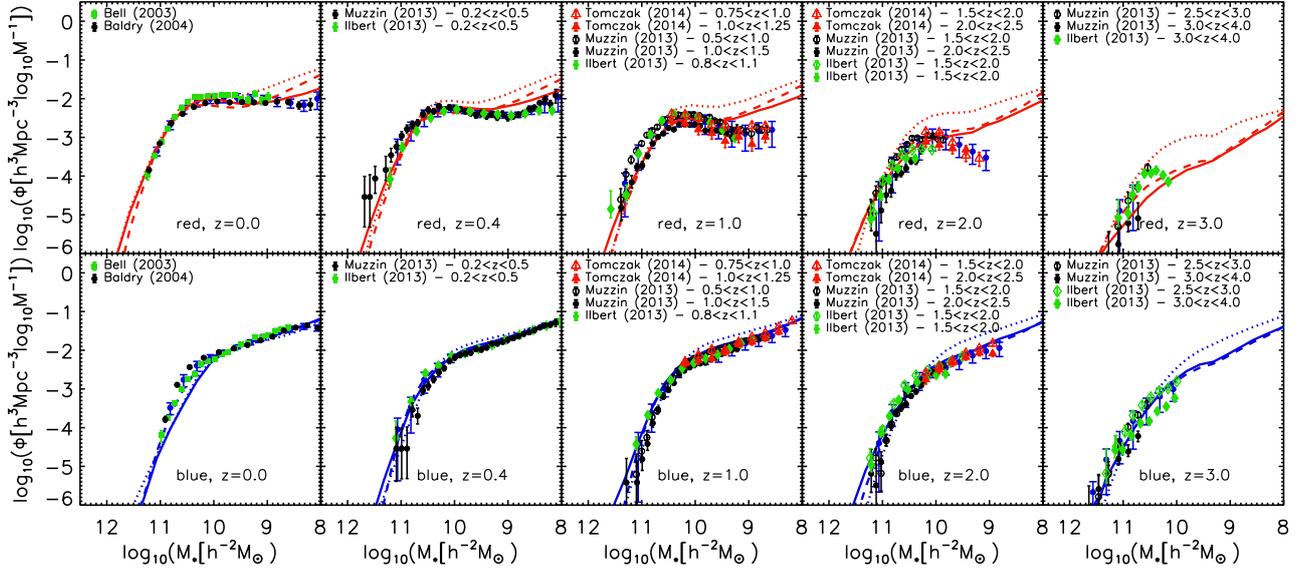

**Figure A2.** Evolution of the stellar mass function of red and blue galaxies from $z = 3$ to 0 as in Fig. 7, but with individual observational data sets shown. These include SDSS data from Bell et al. (2003) and Baldry et al. (2004) at $z = 0$ and ULTRAVISTA (Muzzin et al. 2013, Ilbert et al. 2013) and ZFOURGE (Tomczak et al. 2014) at higher redshifts. The Ilbert et al. (2013) and Tomczak et al. (2014) data have been shifted by -0.14 dex to convert from Bruzual & Charlot (2003) to Maraston (2005) stellar populations (Domínguez Sánchez et al. 2011). For Ilbert et al. (2013) data points this was done even for their lowest redshift bin ($0.2 < z < 0.5$) in order to ensure consistency with the Muzzin et al. (2013) data.

# Galaxy formation in the *Planck* cosmology - I. Supplementary Material

Bruno M. B. Henriques[1], Simon D. M. White[1], Peter A. Thomas[2],
Raul Angulo[3], Qi Guo[4], Gerard Lemson[1], Volker Springel[5,6], Roderik Overzier[7]

[1] *Max-Planck-Institut für Astrophysik, Karl-Schwarzschild-Str. 1, D-85741 Garching b. München, Germany*
[2] *Astronomy Centre, University of Sussex, Falmer, Brighton BN1 9QH, UK*
[3] *Centro de Estudios de Física del Cosmos de Aragón, Plaza San Juan 1, Planta-2, 44001, Teruel, Spain*
[4] *Partner Group of the Max-Planck-Institut für Astrophysik, National Astronomical Observatories, Chinese Academy of Sciences, Beijing, 100012, China*
[5] *Heidelberger Institut für Theoretische Studien, Schloss-Wolfsbrunnenweg 35, D-69118 Heidelberg, Germany*
[6] *Zentrum für Astronomie der Universität Heidelberg, ARI, Mönchhofstr. 12-14, D-69120 Heidelberg, Germany*
[7] *Observatório Nacional/MCTI, Rua José Cristino, 77. CEP 20921-400, São Cristóvão, Rio de Janeiro-RJ, Brazil*

**ABSTRACT**

In this supplementary material we give a full description of the treatment of astrophysical processes in our 2014 model of galaxy formation. This model is built on subhalo merger trees constructed from the Millennium and Millennium-II simulations after scaling to represent the first-year *Planck* cosmology. A set of coupled differential equations allow us to follow the evolution of six baryonic components. Five of these are associated with individual galaxies – a hot gas atmosphere, cold interstellar gas, a reservoir of gas ejected in winds, stars split into bulge, disc and intracluster light components, and central supermassive black holes. The sixth, diffuse primordial gas, is associated with dark matter which is not yet part of any halo. Primordial gas falls with the dark matter onto sufficiently massive haloes, where it is shock-heated. The efficiency of radiative cooling then determines whether it is added directly to the cold gas of the central galaxy, or resides for a while in a hot gas atmosphere. Cold interstellar gas forms stars both quiescently and in merger-induced starbursts which also drive the growth of central supermassive black holes. Stellar evolution not only determines the photometric appearance of the final galaxy, but also heats and enriches its gas components, in many cases driving material into the wind reservoir, from which it may later fall back into the galaxy. Accretion of hot gas onto central black holes gives rise to radio mode feedback, regulating condensation of hot gas onto the galaxy. Environmental processes like tidal and ram-pressure stripping and merging affect the gas components of galaxies, as well as the partition of stars between discs, bulges and the intracluster light, a diffuse component built from tidally disrupted systems. Disc and bulge sizes are estimated from simple energy and angular momentum-based arguments.

**Key words:** galaxies: formation – galaxies: evolution – galaxies: high-redshift – methods: analytical – methods: statistical

## S1 INTRODUCTION

The "Munich" model of galaxy formation is a semi-analytic scheme for simulating the evolution of the galaxy population as a whole and has been continually developed over the last quarter century (White 1989; White & Frenk 1991; Kauffmann et al. 1993, 1999; Springel et al. 2001, 2005). The 2005 completion of the Millennium Simulation enabled implementation of the model on dark matter simulations of high enough resolution to detect the structures associated with the formation of individual galaxies throughout cosmologically relevant volumes. Updates to the baryonic physics have resulted in a series of publicly released galaxy/halo/subhalo catalogues that have been widely used by the community (Croton et al. 2006; De Lucia & Blaizot 2007; Bertone et al. 2007; Guo et al. 2011, 2013).[1] The model of the current paper updates that of Guo et al. (2011), aiming at better representation of the observed build-up over time and of the present star formation activity of the low-mass galaxy population ($8.0 \leqslant \log M_*/\mathrm{M}_\odot \leqslant 9.5$). Guo et al. (2011) itself updated earlier treatments of supernova feedback and of galaxy mergers in order to agree better with observations of dwarf and satellite galaxies. It also introduced detailed tracking of the angular momentum of different galaxy components so that the size evolution of discs and bulges could be followed. Finally, Guo et al. (2013) implemented the procedure of Angulo & White (2010) so that the Millennium Simulation could be used to model evolution in cosmologies other than its native *WMAP1* cosmology.

---

[1] See http://www.mpa-garching.mpg.de/millennium

In this Supplementary Material we aim to give a detailed and fully self-contained summary of the treatment of baryonic physics in our current model. Many aspects of this are unchanged since earlier models but repetition of material in a single coherent and complete description seems preferable to referring each model element back to the particular earlier paper where it was first used. We anticipate updating this supplementary material as future versions of our model are released, so that each will have its own full astrophysics and algorithmic summary.

### S1.1 Dark Matter Simulations

The galaxy formation model of this paper is built on subhalo merger trees describing the evolution of dark matter structures in two large dark matter simulations, the Millennium (Springel et al. 2005) and Millennium-II (Boylan-Kolchin et al. 2009) simulations. Both assume a ΛCDM cosmology with parameters derived by a combined analysis of the 2dFGRS (Colless et al. 2001) and the first-year *WMAP* data (Spergel et al. 2003): $\sigma_8 = 0.9$, $H_0 = 73\,\mathrm{km\,s^{-1}Mpc^{-1}}$, $\Omega_\Lambda = 0.75$, $\Omega_\mathrm{m} = 0.25$, $\Omega_\mathrm{b} = 0.045$ and $n = 1.0$. For this work the original cosmology has been scaled, using the Angulo & White (2010) technique, as updated by Angulo & Hilbert (2015), to represent the best-fitting cosmological parameters derived from the first-year *Planck* data. The underlying cosmology of the dark matter simulations and thus the galaxy formation model is then: $\sigma_8 = 0.829$, $H_0 = 67.3\,\mathrm{km\,s^{-1}Mpc^{-1}}$, $\Omega_\Lambda = 0.685$, $\Omega_\mathrm{m} = 0.315$, $\Omega_\mathrm{b} = 0.0487$ ($f_\mathrm{b} = 0.155$) and $n = 0.96$.

Both the Millennium and Millennium-II simulations trace $2160^3$ (∼10 billion) particles from $z = 127$ to the present day. The Millennium was carried out in a box of original side $500\,h^{-1}\mathrm{Mpc} = 685\,\mathrm{Mpc}$. After rescaling to the *Planck* cosmology, the box size becomes 714 Mpc, implying a particle mass of $1.43 \times 10^9\,\mathrm{M}_\odot$. The Millennium-II follows a region a fifth the linear size, resulting in 125 times better mass resolution. Combined, the two simulations follow dark matter haloes which host galaxies spanning five orders of magnitude in stellar mass at $z = 0$. The particle data were stored in 64 and 68 output snapshots, respectively, for the Millennium and Millennium-II with the last 60 overlapping between the two simulations. After rescaling, the last five snapshots of each simulation correspond to the future, and $z = 0$ corresponds to the sixth from last of the original snapshots. At each time the data were post-processed in order to produce a friend-of-friends (FOF) group catalogue by joining particles separated by less than 20% of the mean interparticle spacing (Davis et al. 1985). The SUBFIND algorithm (Springel et al. 2001) was then applied to identify all the self-bound substructures in each FOF group. The radius of the FOF group is defined as the radius of the largest sphere centered on the potential minimum which contains an overdensity larger than 200 times the critical value. The group mass is then the total mass within this sphere and other group properties are related by:

$$M_\mathrm{200c} = \frac{100}{G} H^2(z) R_\mathrm{200c}^3 = \frac{V_\mathrm{200c}^3}{10\,GH(z)}, \qquad \text{(S1)}$$

where $H(z)$ is the Hubble constant at redshift $z$.

Every subhalo in a given snapshot which contains 20 or more bound particles is connected to a unique descendant in the subsequent snapshot and these links are then used to build subhalo merger trees which encode the assembly history of every subhalo identified at $z = 0$. These trees are the basis on which the galaxy formation model is constructed (see Springel et al. 2005). They allow us to build much more realistic satellite galaxy populations than would be possible using trees linking the FOF haloes themselves. The most massive subhalo in each FOF group is usually much bigger than all the others, and is defined as the "main halo": the group central galaxy (which we often refer to as a "type 0" galaxy) is located at the minimum of the potential of this main halo. All other bound subhaloes contain satellite galaxies at their centres (type 1's). In addition, our galaxy formation model follows satellites which have already lost their own dark matter subhaloes but which are yet to merge with the central galaxy. Such objects are referred to as "type 2" galaxies or "orphan" satellites. Their position and velocity are tied to those of the dark matter particle that was the most bound within their subhalo at the last time that this was identified by SUBFIND with at least 20 particles.

### S1.2 Overview of the galaxy formation physics

Our model for galaxy formation starts by assigning a cosmic abundance of baryons to each collapsed dark matter halo. Subsequent growth brings its fair share of baryons in the form of primordial diffuse gas which shock-heats and then either cools immediately onto the disc of the central galaxy, or is added to a quasi-static hot atmosphere which accretes more slowly through a cooling flow. The disc of cold gas fuels the formation of stars which eventually die, releasing energy, mass and heavy elements into the surrounding medium. This energy reheats cold disc gas, injecting it into a hot atmosphere, which may itself also be ejected into an external reservoir to be reincorporated only at some much later time. Black holes are assumed to grow primarily through the accretion of cold gas during mergers, but also through quiescent accretion from the hot atmosphere, which releases energy which can counteract the cooling flow. This form of feedback eventually curtails star formation in the most massive systems. A number of environmental processes act on satellites as soon as they cross the virial radius of their host. Tidal forces are assumed to remove hot gas, cold gas and stars while hot gas is also removed by ram-pressure stripping. These processes gradually quench star formation, particularly in satellites orbiting within more massive systems. As dark matter subhaloes merge, so do their associated galaxies, although with some delay. Once a subhalo is fully disrupted, its galaxy spirals into the central galaxy, merging after a dynamical friction time and creating a bulge and a burst of star formation. Bulges also form through secular processes whenever discs become sufficiently massive to be dynamically unstable. Finally, the light emitted from stellar populations of different ages is computed via population synthesis models and dust extinction corrections are applied. The uncertain efficiencies and scalings characterising all these physical processes are simultaneously determined by using MCMC techniques to fit a set of calibration observations (in this paper, abundances and passive fractions as a function of stellar mass at a variety of redshifts).

### S1.3 Infall and reionization

Following the standard White & Frenk (1991) approach we assume that each collapsed dark matter structure will, at every time, have a mass of associated baryons given by the cosmic mean baryon fraction, $f_b^{\rm cos} = 15.5\%$ for the *Planck* cosmology. As haloes grow, we assume that matter that was not previously part of any object is added in these same proportions, with the baryons in the form of diffuse primordial gas which shock-heats on accretion, thereafter either cooling again immediately or being added to a quasi-static hot atmosphere.

For sufficiently low-mass haloes and over a large part of cosmic history this simple picture needs modification, since photo-heating by the UV background field raises the temperature of diffuse intergalactic gas to the point where pressure effects prevent it from accreting onto haloes with the dark matter (Efstathiou 1992). In order to model this, we use results from Gnedin (2000) who defines a filtering halo mass, $M_F(z)$, below which the baryonic fraction is reduced with respect to the universal value according to:

$$f_b(z, M_{200c}) = f_b^{\rm cos} \left(1 + (2^{\alpha/3} - 1)\left[\frac{M_{200c}}{M_F(z)}\right]^{-\alpha}\right)^{-3/\alpha}. \quad (S2)$$

For haloes with $M_{200c} > M_F$ suppression of the baryon fraction is small, but for haloes with $M_{200c} \ll M_F(z)$ the baryon fraction drops to $(M_{200c}/M_F(z))^3$. We adopt $\alpha = 2$ and take $M_F(z)$ from the numerical results of Okamoto et al. (2008). $M_F$ varies from $\sim 6.5 \times 10^9 \, {\rm M}_\odot$ at $z = 0$, to $\sim 10^7 \, {\rm M}_\odot$ just before reionization starts at $z = 8$.

### S1.4 Cooling modes

Infalling diffuse gas is expected to shock-heat as it joins a halo. At early times and for low-mass haloes the accretion shock happens close to the central object and the post-shock cooling time is short enough that new material settles onto the cold gas disc at essentially the free-fall rate. At later times and for higher mass haloes the accretion shock moves away from the central object, settling at approximately the virial radius, while the post-shock cooling time exceeds the halo sound crossing time. The shocked heated gas then forms a quasi-static hot atmosphere from which it can gradually accrete to the centre via a cooling flow. The halo mass separating these two regimes is $\sim 10^{12} \, {\rm M}_\odot$ (White & Rees 1978; White & Frenk 1991; Forcada-Miro & White 1997; Birnboim & Dekel 2003). In a realistic, fully three-dimensional situation a hot quasi-static atmosphere can coexist with cold inflowing gas streams in haloes near the transition mass (Kereš et al. 2005; Nelson et al. 2013) but the overall rate of accretion onto the central object remains similar to that given by the formulae below (Benson et al. 2001; Yoshida et al. 2002).

Following the formulation of White & Frenk (1991) and Springel et al. (2001), we assume that, in the quasi-static regime, gas cools from a hot atmosphere where its distribution is isothermal. The cooling time is then given by the ratio between the thermal energy of the gas and its cooling rate per unit volume:

$$t_{\rm cool}(r) = \frac{3\mu m_{\rm H} k T_{200c}}{2\rho_{\rm hot}(r)\Lambda(T_{\rm hot}, Z_{\rm hot})}, \quad (S3)$$

where $\mu m_{\rm H}$ is the mean particle mass, $k$ is the Boltzmann constant, $\rho_{\rm hot}(r)$ is the hot gas density and $Z_{\rm hot}$ is the hot gas metallicity. $T_{\rm hot}$ is the temperature of the hot gas which is assumed to be the virial temperature of the halo given by $T_{200c} = 35.9 \, (V_{200c}/{\rm km\,s^{-1}})^2 \, {\rm K}$ (for subhaloes we use this temperature as estimated at infall). $\Lambda(T_{\rm hot}, Z_{\rm hot})$ is the equilibrium cooling function for collisional processes which depends both on the metallicity and temperature of the gas but ignores radiative ionization effects (Sutherland & Dopita 1993). The hot gas density as a function of radius for a simple isothermal model is given by:

$$\rho_{\rm hot}(r) = \frac{M_{\rm hot}}{4\pi R_{200c} r^2} \quad (S4)$$

and assuming that the cooling radius is where the cooling time equals the halo dynamical time:

$$r_{\rm cool} = \left[\frac{t_{\rm dyn,h} M_{\rm hot} \Lambda(T_{\rm hot}, Z_{\rm hot})}{6\pi \mu m_{\rm H} k T_{200c} R_{200c}}\right]^{\frac{1}{2}}, \quad (S5)$$

where $t_{\rm dyn,h}$ is the halo dynamical time defined as $R_{200c}/V_{200c} = 0.1 H(z)^{-1}$ (De Lucia et al. 2004). The specific choice of coefficient for the dynamical time of the halo is, of course, somewhat arbitrary.

When $r_{\rm cool} < R_{200c}$ we assume that the halo is in the cooling flow regime with gas cooling from the quasi-static hot atmosphere at a rate:

$$\dot{M}_{\rm cool} = M_{\rm hot} \frac{r_{\rm cool}}{R_{200c}} \frac{1}{t_{\rm dyn,h}}. \quad (S6)$$

When $r_{\rm cool} > R_{200c}$ the halo is in the rapid infall regime and material accretes onto the central object in free fall, thus on the halo dynamical time:

$$\dot{M}_{\rm cool} = \frac{M_{\rm hot}}{t_{\rm dyn,h}}. \quad (S7)$$

This particular formula for rapid infall was introduced in Guo et al. (2011) in order to ensure a smooth transition between the two regimes.

### S1.5 Disc formation and angular momentum

As primordial material accretes onto a halo, its dark matter and baryonic components are expected to have similar specific angular momenta. Some of this gas is subsequently added to the central galaxy, and its remaining angular momentum then determines the radius at which it settles within the galactic disc. We follow these processes using the simple model introduced by Guo et al. (2011). The properties of the cold gas and stellar discs are calculated separately and their time evolution is modelled explicitly. The angular

4   *Bruno M. B. Henriques et al.*

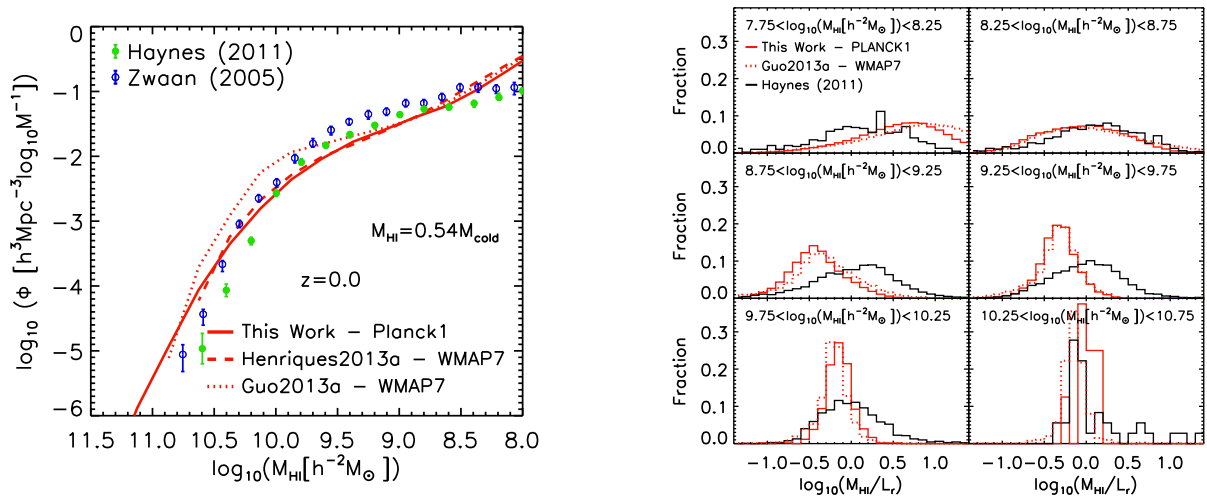

**Figure S1.** Left: The mass function of atomic gas at $z = 0$, defined as $0.54 \times M_{\rm cold}$. Observational data from Zwaan et al. (2005) and Haynes et al. (2011) are compared with the current version of the model (solid red line) and with the models of Henriques et al. (2013) (dashed red line) and Guo et al. (2013) (dotted red line). Right: Histograms of the ratio of cold gas mass to $r-$band luminosity in different bins of atomic gas mass. The current model (solid red line) and that of Guo et al. (2013) (dotted red lines) are compared with the HI-flux-limited sample of Haynes et al. (2011) from ALFALFA (solid black line).

momentum of the cold gas disc changes as a result of star formation and of gas accretion through cooling and minor merger events:

$$\Delta \vec{J}_{\rm gas} = \delta \vec{J}_{\rm gas,cooling} + \delta \vec{J}_{\rm gas,SF} + \delta \vec{J}_{\rm gas,acc}. \quad \text{(S8)}$$

The various terms in this expression can be written out explicitly as:

$$\Delta \vec{J}_{\rm gas} = \frac{\vec{J}_{\rm DM}}{M_{\rm DM}} \dot{M}_{\rm cool} \delta t - \frac{\vec{J}_{\rm gas}}{M_{\rm gas}}((1 - {\rm R}_{\rm ret})\dot{M}_\star \delta t + \Delta M_{\rm reheat})$$
$$+ \frac{\vec{J}_{\rm DM}}{M_{\rm DM}} M_{\rm sat,gas}, \quad \text{(S9)}$$

where $\delta t$ is the time interval considered, $\dot{M}_{\rm cool}$ is the cooling rate (from Eq. S6 or S7), $(1 - {\rm R}_{\rm ret})\dot{M}_\star$ is the formation rate of long lived stars (from eq. S14 below), $\Delta M_{\rm reheat}$ is the cold gas reheated into the hot atmosphere as a result of star formation activity (see eq. S18 and following text below), and $M_{\rm sat,gas}$ is the cold gas mass of any merging satellite(s). Gas coming from cooling events or minor mergers is thus assumed to have the mean specific angular momentum of the dark matter halo ($\vec{J}_{\rm DM}/M_{\rm DM}$), while star formation and reheating are assumed to affect gas with the mean specific angular momentum of the cold gas disc ($\vec{J}_{\rm gas}/M_{\rm gas}$). The stellar disc gains the angular momentum which is removed from the gas disc by star formation events, $\delta \vec{J}_\star = (\vec{J}_{\rm gas}/M_{\rm gas})(1 - {\rm R}_{\rm ret})\dot{M}_\star \delta t$, and in our model, this is the only process that changes the angular momentum of the stellar disc; stars accreted in minor mergers are added to the bulge and secular instabilities are assumed to transfer negligible angular momentum from disc to bulge.

To compute sizes for the stellar and gaseous discs we assume them to have exponential surface density profiles and flat rotation curves. The corresponding scale-lengths can then be calculated as:

$$R_{\rm gas} = \frac{J_{\rm gas}/M_{\rm gas}}{2V_{\rm max}} \quad \text{(S10)}$$

and

$$R_\star = \frac{J_\star/M_{\star,\rm d}}{2V_{\rm max}}, \quad \text{(S11)}$$

where $M_{\rm gas}$ and $M_{\star,\rm d}$ are the total masses of the gas and stellar discs, respectively, and the rotation velocity of both discs is approximated by the maximum circular velocity of their host halo, $V_{\rm max}$. This simple picture needs modification if baryons have a significant impact on the inner structure of their dark matter haloes. We refer the reader to Guo et al. (2011) for further discussion and for a comparison between predicted and observed disc sizes.

With these assumptions, the surface density profiles of the gas and stellar discs are given by:

$$\Sigma_{\rm gas}(R) = \Sigma_{\rm gas,0} \exp(-R/R_{\rm gas}) \quad \text{(S12)}$$

and

$$\Sigma_\star(R) = \Sigma_{\star,0} \exp(-R/R_\star), \quad \text{(S13)}$$

where $\Sigma_{\rm gas,0} = M_{\rm gas}/2\pi R_{\rm gas}^2$ and $\Sigma_{\star,0} = M_\star/2\pi R_\star^2$ are the central surface densities of the cold gas and stellar discs.

### S1.6 Star formation

As noted in the last section, stars are assumed to form from cold gas within the disc of each galaxy. The star formation rate is taken to be:

$$\dot{M}_\star = \alpha_{\rm SF} \frac{(M_{\rm gas} - M_{\rm crit})}{t_{\rm dyn,disk}}, \quad \text{(S14)}$$

where $M_{\rm gas}$ is again the total mass of cold gas, $t_{\rm dyn,disk} = R_\star/V_{\rm max}$, is the dynamical time of the disc, and $M_{\rm crit}$ is a threshold mass (see below). From the total mass of stars formed, $\dot{M}_\star$, we assume that a fraction ${\rm R}_{\rm ret}$ is associated with massive, short-lived, stars and is immediately returned to the cold gas. ${\rm R}_{\rm ret} = 0.43$ is determined from the Chabrier (2003) initial mass function. Thus the stellar mass of the disc is augmented by $\delta M_\star = (1 - R_{\rm ret})\dot{M}_\star \delta t$ and the cold disc mass is reduced by the same amount.

Applying the arguments of Kauffmann (1996) we set the threshold mass for star formation, $M_{\rm crit}$, to be:

$$M_{\rm crit} = M_{\rm crit,0} \left(\frac{V_{\rm 200c}}{200\,{\rm km\,s^{-1}}}\right) \left(\frac{R_{\rm gas}}{10\,{\rm kpc}}\right). \quad (S15)$$

Since Kauffmann et al. (1999) all versions of the Munich model have adopted $M_{\rm crit,0} = 3.8 \times 10^9\,{\rm M_\odot}$ which still appears tenable in comparison with some recent observations in the Milky Way (Lada et al. 2010; Heiderman et al. 2010). However, this and other work (Bigiel et al. 2008; Leroy et al. 2008) suggest that star formation should be linked explicitly to a molecular gas component rather than to the total amount of cold gas. Recently, Fu et al. (2012, 2013) introduced a detailed prescription for the evolution of atomic and molecular components into the Munich semi-analytic model, allowing star formation to be connected directly to molecular content (similar developments where included in the Durham model by Lagos et al. 2011). This is clearly more realistic than eq. (S15) and will be incorporated in future large-scale modelling efforts, but here we simply allow the star formation threshold $M_{\rm crit,0}$ to be a free parameter in our MCMC sampling, recognising that our previous fixed value was poorly justified. The new preferred value is about a factor of two smaller, mainly in order to slow the quenching of satellites by allowing them to use up a larger fraction of the cold gas with which they fall into their host.

Fig. S1 compares our new model's predictions for the atomic gas mass function (left panel) and for the atomic gas over luminosity ratios (right panel) in HI mass-limited bins. The reduced threshold still results in reasonable gas properties but there is a significant deficit of cold gas around the knee of the mass function for atomic gas. In this respect, the earlier model of Guo et al. (2013) does significantly better than the current model, presumably because of its higher star formation threshold and lower star formation efficiency. We note that, since our model does not distinguish between different cold gas phases, an ideal comparison would be with observations of the total cold gas in discs. This would avoid an arbitrary choice for the atomic fraction in the model. However, despite considerable progress in recent years, there are still no large, representative samples providing the HI+H$_2$ content of galaxies. Available H$_2$ samples are limited in terms of both size and selection. The results of these surveys are nevertheless useful, in that they show that the scatter in molecular to atomic ratio is smaller than could be inferred from earlier work (Saintonge et al. 2011, Bothwell et al. 2014), that the typical molecular/total gas fraction is about 25%, dropping to smaller mean values at lower stellar mass, and that few galaxies have a value exceeding 50%. This shows the factor of 0.54 which we use to convert total gas mass to HI mass in Fig.1 (a factor of 0.75 to account for helium and a factor of 0.7 to account for H2) is quite reasonable, and that the scatter in molecular-to-atomic ratio is too small to affect the predicted distributions very strongly, given their width.

Stars can also form whenever two galaxies merge since their cold gas components are strongly disturbed, typically initiating a starburst and feeding some cold gas into the central black hole. This and all other merger-related processes are described in Section S1.12.

### S1.7 Supernova feedback

Massive stars are relatively short-lived. Consequently, soon after an episode of star formation, a large number of them explode as supernovae, strongly clustered both in space and time. The collective energy released by these supernovae and by the stellar winds which precede them is injected into surrounding gas, both cold and hot. As a result, some of the cold interstellar medium is reheated to join the hot gas atmosphere, and this atmosphere itself is also heated, compensating for its cooling and causing some of it to flow out of the galaxy in a wind. This feedback process is a critical aspect of galaxy formation and has long been identified as the main agent controlling its overall efficiency (Larson 1974; White & Rees 1978; Dekel & Silk 1986). As a result, detailed modelling is required if a simulation is to produce a realistic galaxy population. Our specific feedback model is controlled by two main efficiencies, each with three adjustable parameters. One efficiency sets the fraction of the "SN" energy which is available to drive long-term changes in the thermodynamic state of the galaxy's gas components (rather than being lost to cooling radiation), while the other controls the fraction of this energy which is used to reheat cold gas and inject it into the hot gas atmosphere, the remainder being used to heat this atmosphere directly. Heating of the hot atmosphere results in ejection of "wind" material to an external reservoir from which it may or may not be reincorporated at a later time, depending on the mass of the host system.

The energy effectively available to the gas components from supernovae and stellar winds is taken to be:

$$\Delta E_{\rm SN} = \epsilon_{\rm halo} \times \frac{1}{2}\Delta M_\star V_{\rm SN}^2, \quad (S16)$$

where $\frac{1}{2}V_{\rm SN}^2$ is the mean energy injected per unit mass of stars formed (we take $V_{\rm SN} = 630\,{\rm km\,s^{-1}}$) and the efficiency is

$$\epsilon_{\rm halo} = \eta \times \left[0.5 + \left(\frac{V_{\rm max}}{V_{\rm eject}}\right)^{-\beta_2}\right]. \quad (S17)$$

The mass of cold gas reheated by star formation and added to the hot atmosphere is assumed to be directly proportional to the amount of stars formed:

$$\Delta M_{\rm reheat} = \epsilon_{\rm disk}\Delta M_\star, \quad (S18)$$

where the second efficiency is

$$\epsilon_{\rm disk} = \epsilon \times \left[0.5 + \left(\frac{V_{\rm max}}{V_{\rm reheat}}\right)^{-\beta_1}\right]. \quad (S19)$$

This reheating is assumed to require energy $\Delta E_{\rm reheat} = \frac{1}{2}\Delta M_{\rm reheat} V_{\rm 200c}^2$. If $\Delta E_{\rm reheat} > \Delta E_{\rm SN}$, the reheated mass is assumed to saturate at $\Delta M_{\rm reheat} = \Delta E_{\rm SN}/(\frac{1}{2}V_{\rm 200c}^2)$. Otherwise, the remaining SN energy is used to eject a mass $\Delta M_{\rm eject}$ of hot gas into an external reservoir, where

$$\frac{1}{2}\Delta M_{\rm eject} V_{\rm 200c}^2 = \Delta E_{\rm SN} - \Delta E_{\rm reheat}. \quad (S20)$$

There is now considerable observational evidence for ejection of interstellar gas due to star formation activity (Shapley et al. 2003; Rupke et al. 2005; Weiner et al. 2009; Martin et al. 2012; Rubin et al. 2013). While the overall impact of such processes is still debated, observations of rapidly star-forming systems tend to favour mass-loading factors (the ratio of reheated mass to the mass of stars formed) between 1 and 10. The mass-loading factors preferred by our MCMC chains are shown as a function of virial velocity in the top left panel of Fig. S2 and seem similar or somewhat larger than observed.

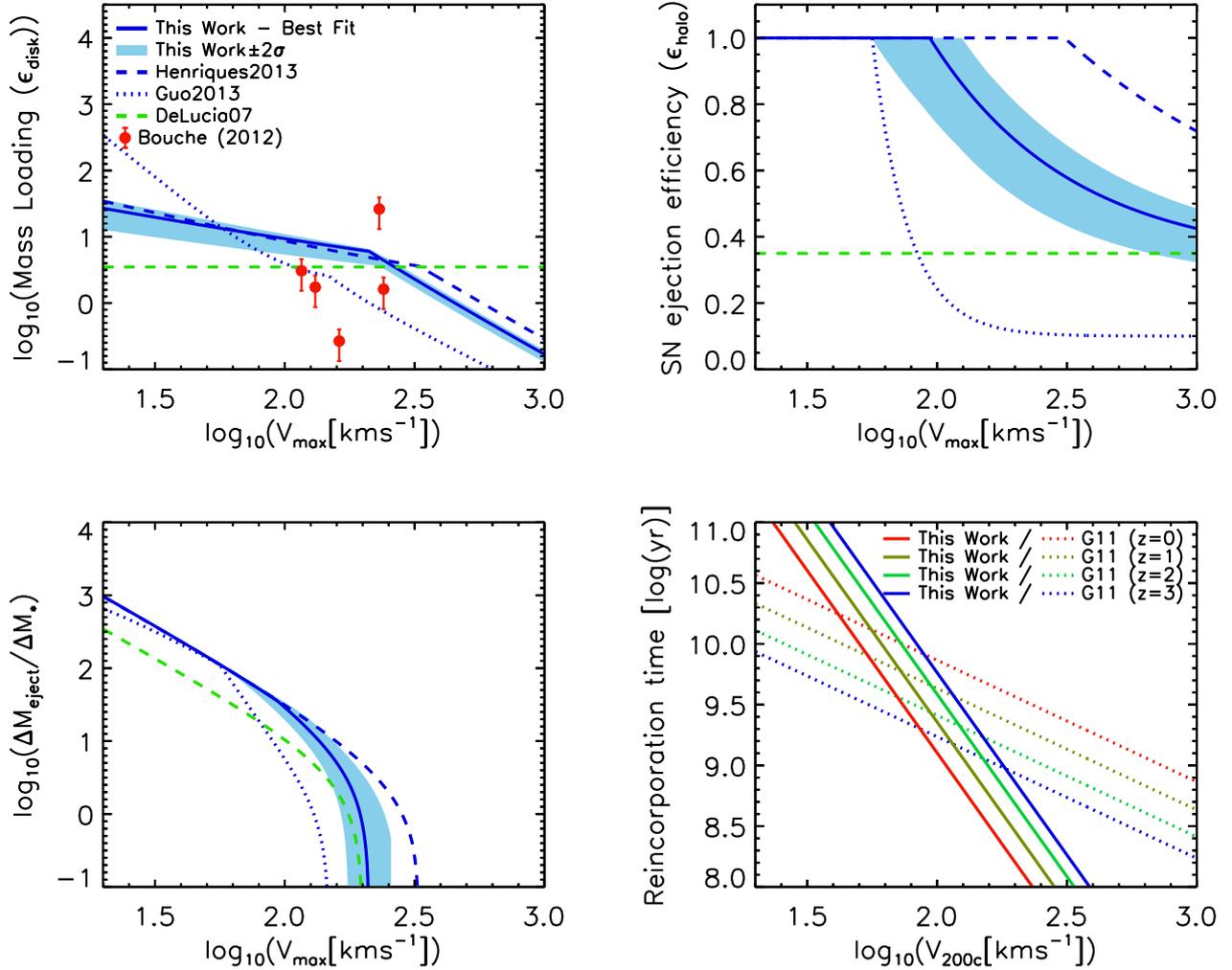

**Figure S2.** Illustration of the dependences of SN feedback on halo properties. The top left panel shows the disc reheating efficiency $\epsilon_{\rm disk}$ as a function of maximum circular velocity $V_{\rm max}$. Often referred to as the mass-loading factor, this is the ratio of the star formation rate to the rate at which ISM material is heated and injected into the hot halo. The top right panel shows the halo ejection efficiency $\epsilon_{\rm halo}$ as a function of $V_{\rm max}$. This is the fraction of the available SN energy which is used in reheating disc gas and in ejecting hot gas from the halo. The bottom left panel gives $\Delta M_{\rm eject}/\Delta M_\star$, the ratio between the hot gas mass ejected to an external reservoir and the cold gas mass which is turned into stars. The bottom right panel shows the reincorporation time-scale $t_{\rm reinc}$ as a function of halo virial velocity $V_{\rm 200c}$ and of redshift (note that the redshift evolution comes solely from the evolution in the relation between $V_{\rm 200c}$ and $M_{\rm 200c}$). In each panel dashed green lines refer to the De Lucia & Blaizot (2007) model, dotted blue lines to the G11-WMAP7 model, dashed blue lines to the Henriques et al. (2013) model and solid blue lines to our new model with its best-fitting parameter values. The blue shaded regions give the $2\sigma$ range allowed by our MCMC sampling. Colours in the bottom right panel indicate redshift as shown by the label.

Ejection of disc gas into the hot atmosphere has relatively little impact when the latter has a short cooling time, since this effectively drives a galactic fountain in which the material soon returns and becomes available for star formation again. Ejection of gas from the hot phase to an external reservoir has substantially stronger long-term effects, however, since such wind ejecta are unavailable for star formation for much longer periods. The top right panel of Fig. S2 shows $\epsilon_{\rm halo}$, the fraction of the available energy that is used in feedback processes, as a function of virial velocity, while the bottom left panel shows $\Delta M_{\rm eject}/\Delta M_\star$, the ratio of the mass of gas ejected in a wind from the galaxy/halo system to the mass of stars formed. For the parameters preferred by our MCMC chains, the available energy is used with high efficiency in low-mass systems, and winds are able to eject material from the haloes of all galaxies with virial velocity less than about $200\,{\rm km\,s^{-1}}$.

### S1.7.1 SN feedback in satellite galaxies [2]

The details of gas reheating and ejection just described, accurately represent the impact of SN feedback in isolated galaxies and in galaxies at the centre of an FoF group (type 0's). For satellites at the centre of a subhalo (type 1's), or orphan satellites with no dark matter, hot or ejected gas (type 2's), the impact of environment must be taken into account. When gas is reheated into the hot phase of a type 1 galaxy, either its own cold gas or gas originated at a type 2 satellite, a fraction is immediately removed due to tidal stripping. In type 2's, reheating will move their own cold gas into the hot phase of their direct central companion, either a type 1 galaxy at the centre of a subhalo or a type 0 galaxy at the centre of the main halo, from which the left-over energy will eject material.

When calculating the amount of energy available from SN and the reheating efficiency, eqs. S17 and S19, we always use the $V_{max}$ of the halo from where the gas will be moved, using the value at infall for satellites. The value for the virial temperature at which reheating saturates and the escape velocity of haloes at which gas is ejected are always taken from the halo where the gas will end up.

## S1.8 Reincorporation of gas ejected in winds

A number of recent papers have argued that most published semi-analytic models and cosmological hydrodynamics simulations form low-mass galaxies ($8.0 \leqslant \log M_*/M_\odot \leqslant 9.5$) too early, leading to an overabundance of these objects at $z \geqslant 1$ (Fontanot et al. 2009; Henriques et al. 2011; Guo et al. 2011; Weinmann et al. 2012; Lu et al. 2014; Genel et al. 2014; Vogelsberger et al. 2014). In the context of the Munich galaxy formation model, the MCMC analysis of Henriques et al. (2013) concluded that this can only be corrected by coupling strong winds in low-mass galaxies with long reincorporation times for the ejecta. This results in slower growth at early times followed by a stronger build-up between $z = 2$ and $0$ as the ejecta finally fall in again.

In the current work we adopt the implementation of Henriques et al. (2013). The mass of gas returned to the hot gas halo from the ejecta reservoir is taken to be:

$$\dot{M}_{\rm ejec} = -\frac{M_{\rm ejec}}{t_{\rm reinc}}, \quad \text{(S21)}$$

where the reincorporation time-scales inversely with the mass of the host halo,

$$t_{\rm reinc} = \gamma' \frac{10^{10}\,\rm M_\odot}{M_{\rm 200c}}, \quad \text{(S22)}$$

rather than with the ratio of its dynamical time and circular velocity, as in Guo et al. (2011). Note that a key aspect of this phenomenological model is that diffuse gas is not available for cooling onto the central galaxy as long as it remains in the external reservoir. The precise location of this reservoir is unspecified, and the gas may not leave the halo entirely. Rather, its entropy may simply be raised above the level assumed by our simple "isothermal" model, in which case the reincorporation time-scales should be interpreted as the time needed to cool to the point where the gas can again be considered part of our standard cooling flow.

---

[2] This section was not included in the published version of the paper, only in the arxiv re-submission.

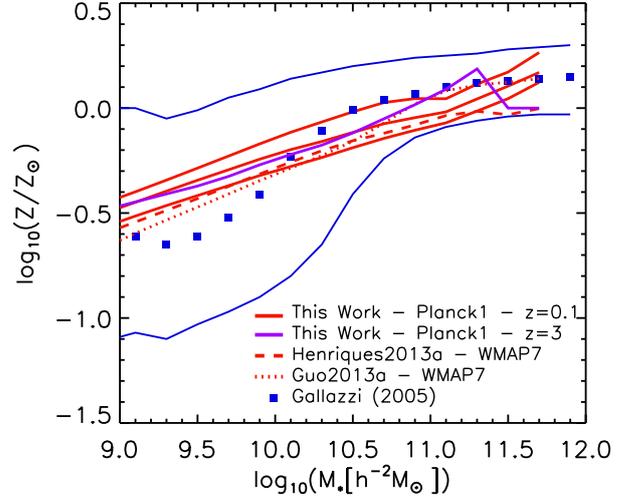

**Figure S3.** Metallicity as a function of stellar mass at $z = 0.1$. Observational data from Gallazzi et al. (2005) are compared to the current model (solid red line) and to the earlier models of Henriques et al. (2013) (dashed red line) and Guo et al. (2013) (dotted red line). Predictions from the current model at $z = 3$ are plotted as a solid purple line.

The differences between the new reincorporation times and those adopted in Guo et al. (2011) are shown as a function of virial velocity and redshift in the lower right panel of Fig. S2. Note that the redshift dependence in our new model simply results from the relation between $M_{\rm 200c}$ and $V_{\rm 200c}$ (eq. S1). In practice gas ejected in winds from low-mass haloes will never be reincorporated unless they become part of a more massive system, while gas returns immediately in the most massive haloes. This implementation agrees qualitatively with the behaviour seen by Oppenheimer & Davé (2008) and Oppenheimer et al. (2010) in their numerical simulations.

## S1.9 Metal enrichment

When stars die, they release newly formed heavy elements into the surrounding medium in addition to mass and energy. In the current work, we follow the total mass of metals only, assuming that each solar mass of stars produces a mass $y$ of heavy metals, with this "yield" treated as a free parameter in the MCMC. The newly formed metals are mixed instantaneously into the cold gas, and thereafter follow it through the various baryonic components of the galaxy, thus enriching the hot gas atmosphere and future generations of stars. In recent work, Yates et al. (2013) and De Lucia et al. (2014) have introduced two different implementations of chemical enrichment into the Munich model. These follow in detail the return of individual elements as stellar populations age, and include metallicity-dependences both in the yields and in population evolution modelling. We expect to incorporate such effects in future large-scale population models, but they are ignored in the model presented here.

The metallicities predicted by the current model are il-

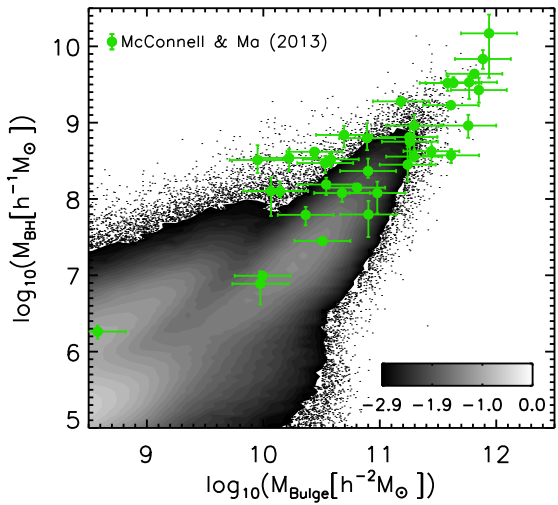

**Figure S4.** Comparison between theoretical predictions from the current model (normalized number density of galaxies shown as logarithmic grey-scale contours and dots) and observations from McConnell & Ma (2013) (green circles) for the black hole-bulge mass relation at $z = 0$

.

lustrated in Fig. S3, where the theoretical stellar mass-stellar metallicity relation at $z = 0$ is compared with observations. The current model and those of Henriques et al. (2013) and Guo et al. (2013) all show similar stellar metallicities, despite significant changes in the treatment of wind ejecta. The $z = 3$ predictions of the current model (and also of the earlier ones) show chemical enrichment to happen very early. The same behaviour is found for the metallicity of the cold gas. This appears to disagree with observation (Maiolino et al. 2008) so further work is clearly needed on this point.

### S1.10  Black hole related processes

In our model, the energy released by supernovae and stellar winds has a dramatic effect on low-mass galaxies, but is unable to reduce cooling onto massive systems ($\log M_*/\,{\rm M}_\odot \geqslant 10.5$) to the very low rates inferred from their observed stellar masses and star formation rates. We follow Croton et al. (2006) in assuming that feedback from central supermassive black holes is the agent that terminates galaxy growth in massive haloes. Black holes are taken to form and to grow when cold gas is driven to the centre of merging systems. In addition, pre-existing black holes merge as soon as their host galaxies do. This "quasar mode" growth is the main channel by which black holes gain mass in our model, but we do not associate it with any feedback beyond that from the strong starbursts which accompany gas-rich mergers. Black holes are also allowed to accrete gas from the hot gas atmospheres of their galaxies, however, and this is assumed to generate jets and bubbles which produce radio mode feedback, suppressing cooling onto the galaxy and so eliminating the supply of cold gas and quenching star formation. The relative importance of these two modes to black hole growth is shown as a function of time and galaxy mass in Fig. 3 of Croton et al. (2006).

#### S1.10.1  Quasar mode - black hole growth

Whenever two galaxies merge, their cold gas components are strongly disturbed and a significant fraction is driven into the inner regions where it may form a black hole or be accreted onto a pre-existing black hole. When both galaxies contain a pre-existing black hole, these are expected to merge during this highly dynamic phase of evolution.

The amount of gas accreted in the quasar mode is taken to depend on the properties of the two merging galaxies as,

$$\Delta M_{\rm BH,Q} = \frac{f_{\rm BH}(M_{\rm sat}/M_{\rm cen})\,M_{\rm cold}}{1 + (V_{\rm BH}/V_{\rm 200c})^2}, \qquad (S23)$$

where $M_{\rm cen}$ and $M_{\rm sat}$ are the total baryon masses of the central galaxy and the satellite which merges with it, $M_{\rm cold}$ is their total cold gas mass, $V_{\rm 200c}$ is the virial velocity of the central halo and $f_{\rm BH}$ and $V_{\rm BH}$ are two adjustable parameters which control the fraction of the available cold gas that is accreted and the virial velocity at which the efficiency saturates. The mass of the black hole at the centre of the final merged galaxy is thus taken to be $M_{\rm BH,f} = M_{\rm BH,1} + M_{\rm BH,2} + \Delta M_{\rm BH,Q}$ where the subscripts 1 and 2 denote the masses of the progenitor black holes.

Mass accretion during mergers is the main channel of black hole growth in our model. The fact that bulges and black holes are formed in related processes results in a tight relation between black hole and bulge masses. The same is seen in observations and is shown in Fig. S4, where we compare the predicted black hole-bulge mass relation for the new model (gray contours) with observations from McConnell & Ma (2013) (green points).

#### S1.10.2  Radio mode - feedback

We assume that central supermassive black holes continually accrete gas from the hot gas atmosphere of their host galaxies, and that this produces radio mode feedback which injects energy into the hot atmosphere. Recent changes to our model have increased the amount of hot gas available to cool onto massive systems at late times, and as a result we find that the original Croton et al. (2006) model for radio mode feedback is unable to suppress star formation sufficiently just above the knee of the galaxy stellar mass function. An MCMC analysis shows that this cannot be solved simply by changing parameters in the original formulation, but that acceptable results can be obtained by assuming the accretion rate to be given by

$$\dot{M}_{\rm BH} = k_{\rm AGN} \left(\frac{M_{\rm hot}}{10^{11}\,{\rm M}_\odot}\right) \left(\frac{M_{\rm BH}}{10^{8}\,{\rm M}_\odot}\right). \qquad (S24)$$

This formula is equivalent to that of Croton et al. (2006) divided by a factor of $H(z)$, so accretion is enhanced at lower redshifts. The differences in the treatment of AGN growth and feedback between the current model and those of Henriques et al. (2013) and Guo et al. (2013) are shown in Fig. S5. Note that in our new model, as in its predecessors, the mass growth of black holes through the radio mode is negligible in comparison with quasar mode accretion.

This form of growth is, however, important in that it is assumed to produce relativistic jets which deposit energy into the hot gas halo in analogy with the hot bubbles seen in galaxy clusters (McNamara & Nulsen 2007; Bîrzan et al.

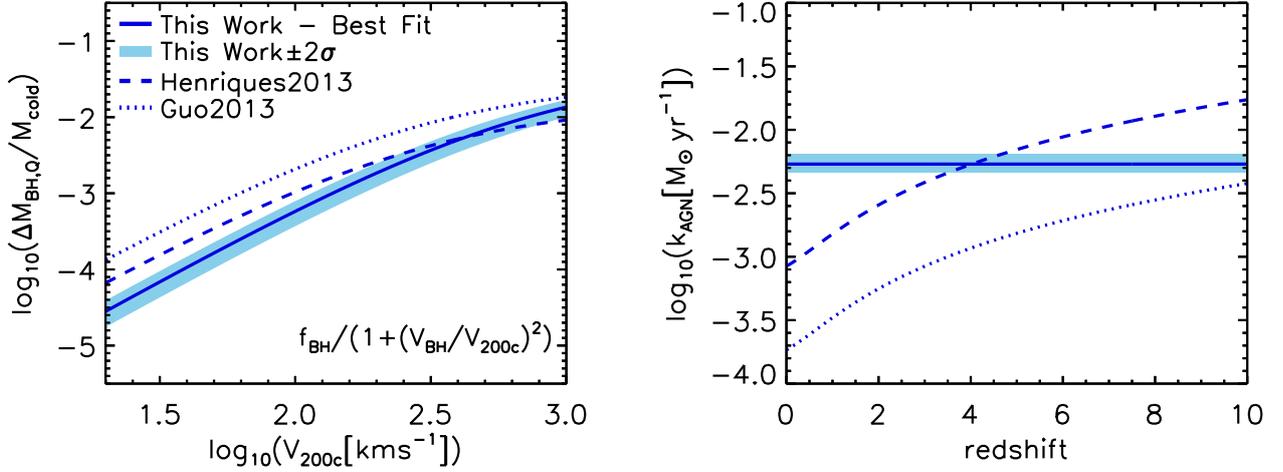

**Figure S5.** The scalings of the processes controlling black hole growth and AGN feedback. The left panel shows the maximum fraction of cold gas accreted (for a major merger of equal mass galaxies) onto central black holes during mergers (quasar accretion) as a function of virial velocity (eq. S23). The right panel shows the ratio of hot gas accretion rate to the product of hot gas and black hole masses (i.e. the coefficient in eq. S24) as a function of redshift. The additional scaling with $V_{200c}^3/M_{200c}$ results in the redshift variation seen in models prior to this work. Accretion in this mode is assumed to suppress cooling in massive systems ($\log M_*/M_\odot \geqslant 10.5$). In both panels the best-fitting and allowed $\pm 2\sigma$ regions for the current model are shown as solid blue lines and light blue regions. The scalings adopted in Henriques et al. (2013) are shown as a dashed blue lines and the Guo et al. (2013) scalings as dotted blue lines.

2004). The energy input rate is taken to be

$$\dot{E}_{\rm radio} = \eta \dot{M}_{\rm BH} c^2, \quad (S25)$$

where $\eta = 0.1$ is an efficiency parameter and $c$ is the speed of light. This energy then suppresses cooling from the hot gas to the cold disc, resulting in an effective cooling rate given by

$$\dot{M}_{\rm cool,eff} = \max\left[\dot{M}_{\rm cool} - 2\dot{E}_{\rm radio}/V_{200c}^2,\ 0\right]. \quad (S26)$$

We assume that elimination of the cooling flow also cuts off the supply of gas to the black hole, so that heating of the hot atmosphere beyond this point is not possible.

Despite growing observational and theoretical evidence for the interaction of black holes with their gaseous environment, we still lack an established theory for this process. The equations given here, like those of Croton et al. (2006), should be regarded as a purely phenomenological representation of some process which acts to prevent the cooling of gas onto massive central galaxies without requiring additional star formation. A good indication that the assumed energetic budget of the process is plausible is given in Fig. S6, where we compare our model predictions to observations of the volume-averaged AGN heating rate. The heating rate in the model is computed as $\Omega_{\rm L\,mech} = \eta \dot{M}_{\rm BH} c^2$, while observationally it is derived from integrating the mechanical luminosity density function. This is obtained from the AGN monochromatic radio to mechanical power (see section 7.3 of Smolčić et al. (2009)). The comparisons with observation presented in the main paper and in Paper II, in combination with Fig. S6, suggest that our current assumptions result in quenching of star formation in intermediate and high-mass galaxies approximately as required by the data.

In addition to the change of eq. S24, a feature that we recently discovered in our dark matter merger trees moti-vated another adjustment to our AGN radio mode feedback. For one massive dark matter structure in the Millennium Simulation, a low-mass satellite subhalo is at some point converted into the main subhalo of the FOF group. As a result, a very small galaxy with a low-mass black hole suddenly acquires $> 10^{14}\,M_\odot$ of hot gas. This leads to a short episode of catastrophic cooling that increases the stellar mass of the central galaxy by as much as two orders of magnitude.

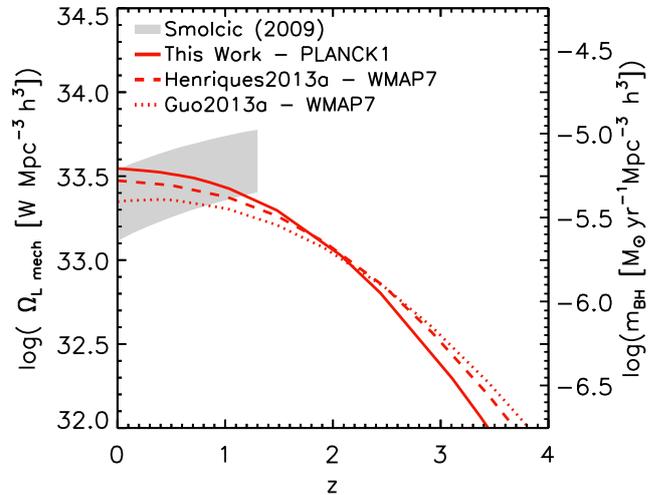

**Figure S6.** Comparison between theoretical predictions and observations for the volume-averaged AGN heating rate across cosmic time. The new model is represented by a solid red line, Henriques et al. (2013) by a dashed red line and Guo et al. (2013) by a dotted red line. Observations from the VLA-COSMOS survey (Smolčić et al. 2009) are shown as a light gray region.

In order to correct this numerical artefact we assume that the AGN energy left over after offsetting the cooling of hot gas in satellite galaxies can be used to offset cooling in the hot gas of their hosts (for satellites within $R_{200}$). As a result, when the low-mass galaxy suddenly becomes the central object of a cluster, the energy released from black holes in other satellite galaxies is enough to suppress excessive cooling from the hot atmosphere onto the central object.

### S1.11 Environmental processes

The growth of structure in a $\Lambda$CDM universe affects galaxies as they and their haloes fall into larger systems and are influenced by tides, by hydrodynamical forces from the hot gas through which they move, and by encounters with other galaxies. Such environmental effects remove material and modify the structure and evolution of the galaxies, in some cases leading to their complete disruption. Several such processes are incorporated in our modelling and their treatment here follows that of Guo et al. (2011) closely. However, environmental effects appear overestimated in the earlier model, which predicts a significantly higher fraction of quenched satellite galaxies than is observed, particularly in intermediate mass haloes ($\log M_{200}/M_\odot \sim 13.0$) (e.g. Wang et al. 2012, 2014). We address this problem here by suppressing ram-pressure stripping in such systems.

#### S1.11.1 Tidal and ram-pressure stripping

As soon a halo falls into a larger system its mass growth reverses as tidal forces begin to remove dark matter (e.g. De Lucia et al. 2004). In the Guo et al. (2011) model, this implies that no new baryonic material is added to the system and its hot gas atmosphere is stripped away in proportion to its dark matter mass,

$$\frac{M_{\rm hot}(R_{\rm tidal})}{M_{\rm hot,infall}} = \frac{M_{\rm DM}}{M_{\rm DM,infall}}, \quad (S27)$$

where the limiting radius is given by a simple "isothermal" model,

$$R_{\rm tidal} = \left(\frac{M_{\rm DM}}{M_{\rm DM,infall}}\right) R_{\rm DM,infall}. \quad (S28)$$

In these equations, $M_{\rm DM,infall}$, $R_{\rm DM,infall}$ and $M_{\rm hot,infall}$ are $M_{\rm 200c}$, $R_{\rm 200c}$ and the hot gas mass of the halo just prior to infall, and $M_{\rm DM}$ and $M_{\rm hot}$ are the current masses of these two components. By construction, tidal stripping will have removed all hot gas once the subhalo is disrupted and the galaxy becomes an orphan.

Hot gas can also be stripped by ram-pressure effects which are followed starting when the satellite first falls within the virial radius of its host. At a certain distance $R_{\rm r.p.}$ from the centre of the satellite, self-gravity is approximately balanced by ram pressure:

$$\rho_{\rm sat}\left(R_{\rm r.p.}\right) V_{\rm sat}^2 = \rho_{\rm par}\left(R\right) V_{\rm orbit}^2, \quad (S29)$$

where $\rho_{\rm sat}(R_{\rm r.p.})$ is the hot gas density of the satellite at radius $R_{\rm r.p.}$, $V_{\rm sat}$ is the virial velocity of the subhalo at infall (which we assume to be constant as the subhalo orbits around the main halo), $\rho_{\rm par}(R)$ is the hot gas density of the parent dark matter halo at the distance $R$ of the satellite from the centre of its potential well, and $V_{\rm orbit}$ is the orbital velocity of the satellite, which we approximate as the virial circular velocity of the main halo. The densities here are again estimated from the total mass and limiting radius of the relevant component according to an "isothermal" model, $\rho \propto r^{-2}$. Finally, the radius of the hot gas component is taken to be the smaller of $R_{\rm r.p.}$ and $R_{\rm tidal}$.

In the current work we apply this ram-pressure model only in haloes above a threshold mass ($M_{\rm r.p.}$) which we introduce as a free parameter which observational constraints then require to be $\sim 10^{14}\,{\rm M_\odot}$. Combined with our lower threshold for star formation, this changes reduces the excess of passive satellites found in the Guo et al. (2011) and Henriques et al. (2013) models, while remaining consistent with observation of ram-pressure stripping phenomena in rich clusters.

Finally we note that ram-pressure effects on the cold gas component are not included in our model. Such effects are expected (e.g. Bekki 2014) and are indeed observed in high density regions (e.g. Crowl et al. 2005; Fumagalli et al. 2014) but they require more extreme conditions than the effects considered in this section.

#### S1.11.2 Tidal disruption of galaxies

Our implementation of the tidal disruption of the stellar and cold gas components of galaxies is unchanged from Guo et al. (2011). Since both components are considerably more concentrated than the dark matter, we consider disruption only for galaxies that have already lost their dark matter and hot gas components. For such orphans, the baryonic (cold gas + stellar mass) density within the half-mass radius is compared to the dark matter density of the main halo within the pericentre of the satellite's orbit. If the latter is larger, i.e.

$$\frac{M_{\rm DM,halo}(R_{\rm peri})}{R_{\rm peri}^3} \equiv \rho_{\rm DM,halo} > \rho_{\rm sat} \equiv \frac{M_{\rm sat}}{R_{\rm sat,half}^3}, \quad (S30)$$

the satellite is completely disrupted, its stars are added to the intracluster light (ICL) and its cold gas is added to the hot gas atmosphere of the central galaxy. The galaxy's half-mass radius is calculated from those of the cold gas and stellar discs and the bulge (assuming exponential surface density profiles for the first two and a surface density scaling with $r^{1/4}$ for the latter), while its orbital pericentre is calculated as

$$\left(\frac{R}{R_{\rm peri}}\right)^2 = \frac{\ln R/R_{\rm peri} + \frac{1}{2}(V/V_{\rm 200c})^2}{\frac{1}{2}(V_t/V_{\rm 200c})^2}, \quad (S31)$$

assuming conservation of energy and angular momentum and a singular isothermal potential for the orbit, $\phi(R) = V_{\rm 200c}^2 \ln R$. In these equations, $R$ is the current distance of the satellite from halo centre, and $V$ and $V_t$ are the total and tangential velocities of the satellite with respect to halo centre (see Section S1.12.1 for a description on how these are determined for orphans). We tested that this condition for complete disruption of satellites gives very similar answers to the more detailed implementation of gradual stripping proposed by Henriques & Thomas (2010) (See Contini et al. (2014) for a more extensive comparison of different implementations of tidal disruption).

*S1.11.3 SN feedback in orphan galaxies*

For orphan galaxies environmental effects are particularly dramatic. Since our implementation of tidal stripping of hot gas is directly connected to the stripping of dark matter, once galaxies lose their halo, they also have no hot gas left. As described in Section S1.7.1, from this point on, we also assume that any cold gas reheated by star formation activity leaves the galaxy and is added to the hot gas atmosphere of the main halo. This can lead to rapid depletion of any remaining cold gas.

**S1.12 Mergers and bulge formation**

*S1.12.1 Positions and velocities of orphans*

Once a satellite subhalo is disrupted, its central galaxy becomes an orphan and its position and velocity are linked to those of the dark matter particle which was most strongly bound within the subhalo just prior to its disruption. As soon as a disruption event occurs, this particle is identified and a merging clock is started, based on an estimate of how long the satellite will take to spiral into the central object due to dynamical friction. This time is computed using the Binney & Tremaine (1987) formula:

$$t_{\rm friction} = \alpha_{\rm friction} \frac{V_{200c} r_{\rm sat}^2}{G M_{\rm sat} \ln\Lambda}, \quad (S32)$$

where $M_{\rm sat}$ is the total mass of the satellite (dark and baryonic), $\ln\Lambda = \ln(1 + M_{200c}/M_{\rm sat})$ is the Coulomb logarithm and $\alpha_{\rm frction} = 2.4$ is a parameter originally set by De Lucia & Blaizot (2007) to match the bright end of the $z = 0$ luminosity functions. This value was later shown to be consistent with inferences from direct numerical simulation (Boylan-Kolchin et al. 2008; De Lucia et al. 2010) but should still be considered poorly known. The Millennium-II simulation is able to resolve subhaloes which have been stripped to masses below that of their central galaxy. In such cases we turn on the merging clock as soon as the subhalo mass drops below the stellar mass in the galaxy.

Following Guo et al. (2011) we model the decay of the satellite's orbit due to dynamical friction by placing the orphan galaxy not at the current position of the particle with which it is identified, but at a position whose (vector) offset from the central galaxy is reduced from that of the particle by a factor of $(1-\Delta t/t_{\rm friction})$ where $\Delta t$ is the time since the dynamical friction clock was started. The (vector) velocity of the orphan galaxy is set equal to that of the tagged particle. This time dependence is based on a simple model for a satellite with "isothermal" density structure spiralling to the centre of an isothermal host on a circular orbit. When $\Delta t = t_{\rm friction}$ the orphan merges with the central galaxy.

*S1.12.2 Merger-triggered star formation*

When a satellite finally merges with the object at the centre of the main halo, the outcome is different for major and minor mergers. We define a major merger to be one in which the total baryonic mass of the less massive galaxy exceeds a fraction $R_{\rm merge}$ of that of the more massive galaxy. In a major merger, the discs of the two progenitors are destroyed and all their stars become part of of the bulge of the descendent, along with any stars formed during the merger. In a minor merger, the disc of the larger progenitor survives and accretes the cold gas component of the smaller galaxy, while its bulge accretes all the stars of the victim. Stars formed during the merger stay in the disc of the descendent. In both types of merger, cold gas is fed to the central black hole according to the formulae of Section S1.10.1. $R_{\text{merge}}$ is fixed and not included in the MCMC analysis since we find some tension between the values required for model predictions to reproduce observed colours and observed morphologies. Our adopted value of 0.1, slightly compromises the agreement with the observed red fraction of galaxies at $z = 1$, but ensures that the morphologies of massive galaxies in the model are reasonably close to observations. This comparison is shown in Fig. S7. Some discrepancies remain at low mass, particularly for MRII, where there is an excess of normal spirals with respect to pure disc galaxies.

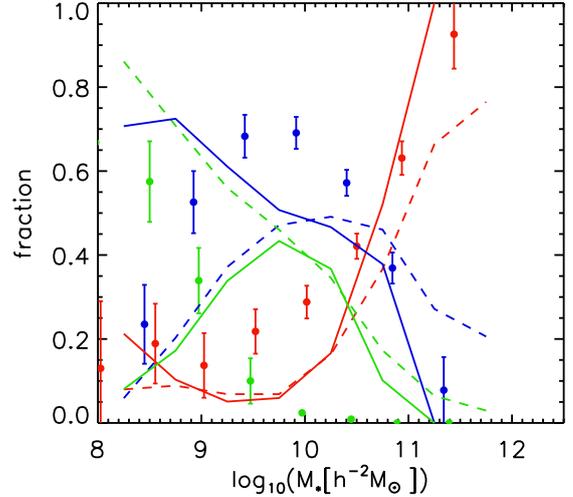

**Figure S7.** Comparison between theoretical predictions and observations for the fraction of different morphological types as a function of stellar mass. Red lines show the fraction of ellipticals ($M_{\rm bulge}/M_{\rm total} > 0.7$), blue lines show the fraction of normal spirals ( $0.01 < M_{\rm bulge}/M_{\rm total} < 0.7$) and green lines represent pure discs or extreme late-types ($M_{\rm bulge}/M_{\rm total} < 0.01$). Solid and dashed lines indicate predictions based on the MRII and MR simulations, respectively. Observational data points from Conselice (2006) are represented by the filled circles.

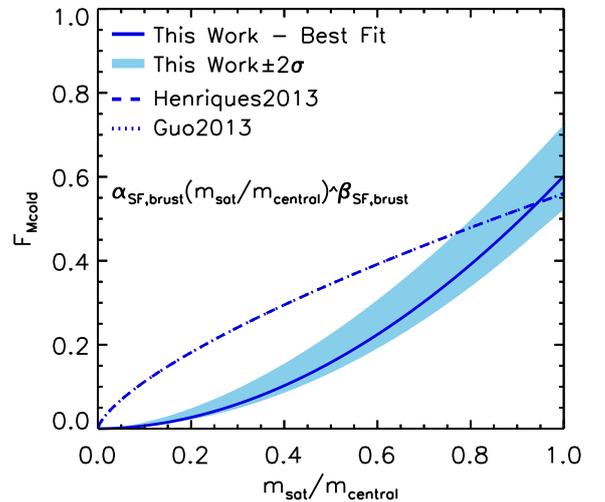

**Figure S8.** The fraction of cold gas converted to stars in merger-triggered starbursts as a function of the mass ratio between satellite and central galaxies. The best-fitting and $\pm 2\sigma$ regions for the current model are shown as a solid blue line and a light blue region respectively. The scaling adopted in Henriques et al. (2013) and Guo et al. (2013) is shown as a dashed blue line.

The stellar mass formed during a merger is modelled using the "collisional starburst" formulation of Somerville et al. (2001):

$$M_{\star,\text{burst}} = \alpha_{\text{SF,burst}} \left(\frac{M_1}{M_2}\right)^{\beta_{\text{SF,burst}}} M_{\text{cold}}, \quad (S33)$$

where $M_1 < M_2$ are the baryonic masses of the two galaxies, and $M_{\text{cold}}$ is their total cold gas mass. The $\alpha_{\text{SF,burst}}$ and $\beta_{\text{SF,burst}}$ parameters were originally fixed to reflect the results of the Mihos & Hernquist (1996) simulations, but in the current work they are left free and are allowed to vary in our MCMC analysis. Despite this, in our best-fitting model the fraction of cold gas converted to stars in merger-related bursts is relatively close to what was previously assumed. Fig. S8 compares this quantity for our current model (solid blue line and light blue regions) to that assumed in the models of Henriques et al. (2013) and Guo et al. (2013) (dashed blue line).

*S1.12.3  Bulge Formation*

In our model bulges can form through major and minor mergers and through the buckling instability of discs. After a major merger, all stars are considered part of the new bulge, but the remnant of a minor merger retains the stellar disc of the larger progenitor and its bulge gains only the stars from the smaller progenitor. Following Guo et al. (2011), we use energy conservation and the virial theorem to compute the change in sizes in both minor and major mergers:

$$\frac{GM_{\text{new,bulge}}^2}{R_{\text{new,bulge}}} = \frac{GM_1^2}{R_1} + \frac{GM_2^2}{R_2} + 2\alpha_{\text{inter}}\frac{GM_1M_2}{R_1 + R_2}, \quad (S34)$$

where the left-hand side represents the binding energy of the final bulge, the first two terms of the right-hand side represent the binding energies of the progenitor stellar systems (the radii in these three terms are taken to be the half-mass radii of the corresponding stellar systems) and the last term is the binding energy of the relative orbit of the two progenitors at the time of merger. The coefficient $\alpha_{\text{inter}}$ quantifies the binding energy invested in this orbit relative to that in the individual systems. Guo et al. (2011) set $\alpha_{\text{inter}} = 0.5$ and show that this leads to bulge sizes in reasonable agreement with SDSS data. When either of the progenitors is a composite disc+bulge system, its half-mass radius is calculated assuming an exponential disc and an $r^{1/4}$-law bulge.

Another important channel of bulge growth is secular evolution through disc instabilities. These dynamical instabilities occur through the formation of bars which then buckle. They transport material inwards to the bulge and they occur in galaxies where self-gravity of the disc dominates the gravitational effects of the bulge and halo. As a criterion for disc instability, we follow Guo et al. (2011) in adopting

$$V_{\text{max}} < \sqrt{\frac{GM_{\star,\text{d}}}{3R_{\star,\text{d}}}}, \quad (S35)$$

where $M_{\star,\text{d}}$ and $R_{\star,\text{d}}$ are the stellar mass and exponential scale-length of the stellar disc and $V_{\text{max}}$ is the maximum circular velocity of the host dark matter halo hosting the disc.

When the instability criterion of eq. S35 is met, we transfer sufficient stellar mass from the disc to the bulge to make the disc marginally stable again. Following Guo et al. (2011) we assume that this mass $\delta M_\star$ is transferred from the innermost part of the disc. Thus, the half-mass radius of the material to be added to the bulge, $R_{\text{b}}$, is related to $\delta M_\star$ through

$$\delta M_\star = 2\pi\Sigma_{\star 0}R_{\star,d}[R_{\star,d} - (R_{\text{b}} + R_{\star,d})\exp(-R_{\text{b}}/R_{\star,d})], \quad (S36)$$

where $\Sigma_{\star 0}$ and $R_{\star,d}$ are the central surface density and exponential scale-length of the unstable disc. We neglect the angular momentum of the transferred material, so the disc's angular momentum is unchanged, resulting in an increase in $R_{\star,d}$ to compensate for the mass lost.

If the galaxy already has a spheroidal component, the newly created bulge material is assumed to merge with the existing bulge according to eq. (S34) where we now take $\alpha_{\text{inter}} = 2$ to account for the fact that the inner disc and the initial bulge are concentric and have no relative motion. For further discussion of this model for bulge growth and for comparisons with observational data we refer the reader to Guo et al. (2011).

**S1.13  Stellar populations synthesis**

Stellar population synthesis models are a crucial part of galaxy formation theory as they link the masses, ages and metallicities predicted for stars to the observable emission at various wavelengths. We use Maraston (2005) as our default stellar population synthesis model, but we have checked that the publicly released but still unpublished Charlot & Bruzual (2007) code leads to very similar results for all the properties we consider. Somewhat different predictions are obtained with the earlier Bruzual & Charlot (2003) code because of the weaker emission it assumes for the TP-AGB stage of evolution of intermediate age stars. Recent work by a number of authors suggests that the more recent models are in better agreement with observed near-infrared emission from bright galaxies at $z \geqslant 2$ (Henriques et al. 2011, 2012; Tonini et al. 2009, 2010; Fontanot & Monaco 2010; Tonini et al. 2011; Gonzalez-Perez et al. 2014). For the Munich galaxy formation model, in particular, Henriques et al. (2011) and Henriques et al. (2012) showed that Maraston (2005) or Charlot & Bruzual (2007) populations give stellar mass and $K$-band luminosity functions for which the massive/bright end agrees with observation from $z = 3$ to 0.

Nevertheless, as part of our model release we will, for comparison purposes, also include luminosities computed using Bruzual & Charlot (2003) stellar populations.

### S1.14 Dust Model

Actively star-forming galaxies are known to be rich in dust. This can have a dramatic effect on their emitted spectrum since dust significantly absorbs optical/UV light while having a much milder effect at longer wavelengths. As a result, dust-dominated galaxies will generally have red colours even if they are strongly star forming. We follow De Lucia & Blaizot (2007) in considering dust extinction separately for the diffuse interstellar medium (ISM) (following Devriendt et al. 1999) and for the molecular birth clouds within which stars form (following Charlot & Fall 2000)). The optical depth of dust as a function of wavelength is computed separately for each component and then a slab geometry is assumed in order to compute the total extinction of the relevant populations. We do not at present attempt to compute the detailed properties of the dust particles or the re-emission of the absorbed light.

#### S1.14.1 Extinction by the ISM

The optical depth of diffuse dust in galactic discs is assumed to vary with wavelength as

$$\tau_\lambda^{ISM} = (1+z)^{-1} \left(\frac{A_\lambda}{A_V}\right)_{Z_\odot} \left(\frac{Z_{\rm gas}}{Z_\odot}\right)^s$$
$$\times \left(\frac{\langle N_H \rangle}{2.1 \times 10^{21}\,{\rm atoms\,cm}^{-2}}\right), \quad (S37)$$

where $\langle N_H \rangle$ represents the mean column density of hydrogen and is given by

$$\langle N_H \rangle = \frac{M_{\rm cold}}{1.4\, m_p \pi (a R_{\rm gas,d})^2}. \quad (S38)$$

Here $R_{\rm gas,d}$ is the cold gas disc scale-length, 1.4 accounts for the presence of helium and $a = 1.68$ in order for $\langle N_H \rangle$ to represent the mass-weighted average column density of an exponential disc. Following the results in Guiderdoni & Rocca-Volmerange (1987), the extinction curve in eq. (S37) depends on the gas metallicity and is based on an interpolation between the Solar Neighbourhood and the Large and Small Magellanic Clouds: $s = 1.35$ for $\lambda < 2000$ Å and $s = 1.6$ for $\lambda > 2000$ Å. The extinction curve for solar metallicity, $(A_\lambda/A_V)_{Z_\odot}$, is taken from Mathis et al. (1983).

The redshift dependence in eq. (S37) is significantly stronger than in previous versions of our model ($(1+z)^{-0.5}$ in Kitzbichler & White (2007) and $(1+z)^{-0.4}$ in Guo & White (2009). The dependence implies that for the same amount of cold gas and the same metal abundance, there is less dust at high redshift. The motivation comes both from observations (Steidel et al. 2004; Quadri et al. 2008) and from the assumption that dust is produced by relatively long-lived stars. However, it may also be that this redshift dependence has to be introduced as a phenomenological compensation for the excessively early build-up of the metal content in model galaxies shown in Fig. S3. In practice we include it merely as representation of the observed extinction behaviour (Bouwens et al. 2014), while the physical reason for discrepancy with the simple model remains to be established. As will be shown in Clay et al. (2015) this produces luminosity functions and extinction estimates for Lyman-break galaxies at $z > 5$ compatible with HST data.

#### S1.14.2 Extinction by molecular clouds

This second source of extinction affects only young stars. Following Charlot & Fall (2000), our model assumes that such extinction affects stars younger than the lifetime of stellar birth clouds (taken to be $10^7$ years). The relevant optical depth is taken to be

$$\tau_\lambda^{BC} = \tau_\lambda^{\rm ISM} \left(\frac{1}{\mu} - 1\right) \left(\frac{\lambda}{5500\text{Å}}\right)^{-0.7}, \quad (S39)$$

where $\mu$ is given by a random Gaussian deviate with mean 0.3 and standard deviation 0.2, truncated at 0.1 and 1.

#### S1.14.3 Overall extinction curve

In order to get the final overall extinction, every galaxy is assigned an inclination given by the angle between the disc angular momentum and the $z$-direction of the simulation box, and a "slab" geometry is assumed for the dust in the diffuse ISM. For sources that are uniformly distributed within the disc then the mean absorption coefficient is

$$A_\lambda^{\rm ISM} = -2.5 \log_{10}\left(\frac{1 - \exp^{-\tau_\lambda^{\rm ISM}\sec\theta}}{\tau_\lambda^{\rm ISM}\sec\theta}\right), \quad (S40)$$

where $\theta$ is the angle of inclination of the galaxy relative to the line-of-sight. Emission from young stars embedded within birth clouds is subject to an additional extinction of

$$A_\lambda^{\rm BC} = -2.5 \log_{10}\left(\exp^{-\tau_\lambda^{\rm BC}}\right). \quad (S41)$$

### S1.15 Monte Carlo Markov Chains

In order to sample the full multidimensional parameter space of our model we use MCMC techniques. This enables exploration of the allowed regions when the model is constrained by a broad variety of calibrating observations, which may be of different types and correspond to different redshifts. The same scheme allows us to assess the merits of different implementations of critical astrophysical processes. We use a version of the Metropolis-Hastings method (Metropolis et al. 1953; Hastings 1970); a full description of the algorithm can be found in Section 3 of Henriques et al. (2009). A full MCMC chain requires evaluation of many tens of thousands of models and it is not computationally feasible to build all of these models for the full Millennium or Millennium-II simulation. We therefore use sampling techniques to construct a representative subset of subhalo merger trees on which the galaxy formation model is evaluated during the MCMC procedure (details are given in Appendix 2 of Henriques et al. 2013). Once the best-fitting model has been identified, it can be implemented on the full volumes of the two simulations.

Fig. S9 shows marginalized 1D posterior distributions for our model parameters when the model is constrained by observational data on the abundance and passive fraction of galaxies as a function of stellar mass from $z = 3$ down to $z = 0$. Vertical solid blue lines correspond to the parameter values of the best-fitting model (taken to be the one for which the MCMC chain found the highest likelihood) and these are also presented in Table S1. Shaded blue regions show the central 95% confidence region of each marginalized posterior distribution and the boundaries of the corresponding parameter interval are also given in Table S1. Interestingly, although the best-fitting model has parameters which lie within these regions in almost all cases, this is not true for $k_{\rm AGN}$ and $V_{\rm reheat}$. The allowed parameter range is quite narrow in all cases, showing that these observations are sufficient to specify our model completely without major degeneracies.

**Table S1.** Results from the MCMC parameter estimation. The best-fitting values of parameters and their "2-$\sigma$" confidence limits are compared with the values published in Henriques et al. (2013) and Guo et al. (2013).

| | Guo13 | Henriques13 | New Best Fit | 2$\sigma$ lower | 2$\sigma$ upper | Units |
|---|---|---|---|---|---|---|
| $\alpha_{\rm SF}$ (SF eff - eq. S14) | 0.011 | 0.055 | 0.025 | 0.021 | 0.029 | |
| $M_{\rm crit,0}$ (Gas mass threshold - eq. S15) | 0.38 | 0.38 | 0.24 | 0.20 | 0.27 | $10^{10}\,{\rm M}_\odot$ |
| $\alpha_{\rm SF,burst}$ (SF burst eff - eq. S33) | 0.56 | 0.56 | 0.60 | 0.52 | 0.73 | |
| $\beta_{\rm SF,burst}$ (SF burst slope - eq. S33) | 0.70 | 0.70 | 1.9 | 1.7 | 2.0 | |
| $k_{\rm AGN}$ (Radio feedback eff - eq. S24) | new eq | new eq | $5.3 \times 10^{-3}$ | $4.6 \times 10^{-3}$ | $6.5 \times 10^{-3}$ | ${\rm M}_\odot\,{\rm yr}^{-1}$ |
| $f_{\rm BH}$ (BH growth eff - eq. S23) | 0.03 | 0.015 | 0.041 | 0.035 | 0.048 | |
| $V_{\rm BH}$ (Quasar growth scale - eq. S23) | 280 | 280 | 750 | 670 | 880 | ${\rm km\,s}^{-1}$ |
| $\epsilon$ (Mass-loading eff - eq. S19) | 4.0 | 2.1 | 2.6 | 1.9 | 2.6 | |
| $V_{\rm reheat}$ (Mass-loading scale - eq. S19) | 80 | 405 | 480 | 390 | 540 | ${\rm km\,s}^{-1}$ |
| $\beta_1$ (Mass-loading slope - eq. S19) | 3.2 | 0.92 | 0.72 | 0.60 | 0.82 | |
| $\eta$ (SN ejection eff - eq. S17) | 0.18 | 0.65 | 0.62 | 0.53 | 0.68 | |
| $V_{\rm eject}$ (SN ejection scale - eq. S17) | 90 | 336 | 100 | 90 | 120 | ${\rm km\,s}^{-1}$ |
| $\beta_2$ (SN ejection slope - eq. S17) | 3.2 | 0.46 | 0.80 | 0.71 | 0.95 | |
| $\gamma$ (Ejecta reincorporation - eq. S22) | new eq | $1.8 \times 10^{10}$ | $3.0 \times 10^{10}$ | $2.6 \times 10^{10}$ | $3.6 \times 10^{10}$ | yr |
| $M_{\rm r.p.}$ (Ram-pressure threshold) | 0.0 | 0.0 | $1.2 \times 10^{4}$ | $1.1 \times 10^{4}$ | $1.6 \times 10^{4}$ | $10^{10}\,{\rm M}_\odot$ |
| $R_{\rm merger}$ (Major-merger threshold) | 0.30 | 0.30 | 0.1 | | | |
| $\alpha_{\rm friction}$ (Dynamical friction - eq. S32) | 2.0 | 2.0 | 2.5 | 2.1 | 2.8 | |
| $y$ (Metal yield) | 0.03 | 0.047 | 0.046 | 0.038 | 0.053 | |

Fig. S9 also shows the parameter values corresponding to the best-fitting models of Henriques et al. (2013) (dashed vertical blue lines) and Guo et al. (2013) (dotted vertical blue lines). Despite changes in cosmology and in several aspects of the astrophysical modelling, the efficiencies of most processes are very similar in the different versions of the model (see also Figs S2, S5 and S8). This indicates that parameters which were not previously included in the MCMC sampling (all of them in the case of Guo et al. (2013)!) were, in fact, well constrained by less rigorous comparison to observations. The exceptions are the cold gas density threshold for star formation, which now has a significantly lower value, and the ram-pressure stripping threshold, which was zero in the earlier models. Both changes are required to predict the correct evolution of the fraction of passive galaxies as a function of stellar mass, which simply could not be explained by the previous models. Indeed, if we were to carry out an MCMC analysis of the Henriques et al. (2013) or the Guo et al. (2013) model using this observational constraint, we would find a very low maximum likelihood value. As a result, we do not need to integrate over the posterior distributions to perform Bayesian model selection – the current model is the only one of the three which can come close to representing the observational distributions used as constraints.

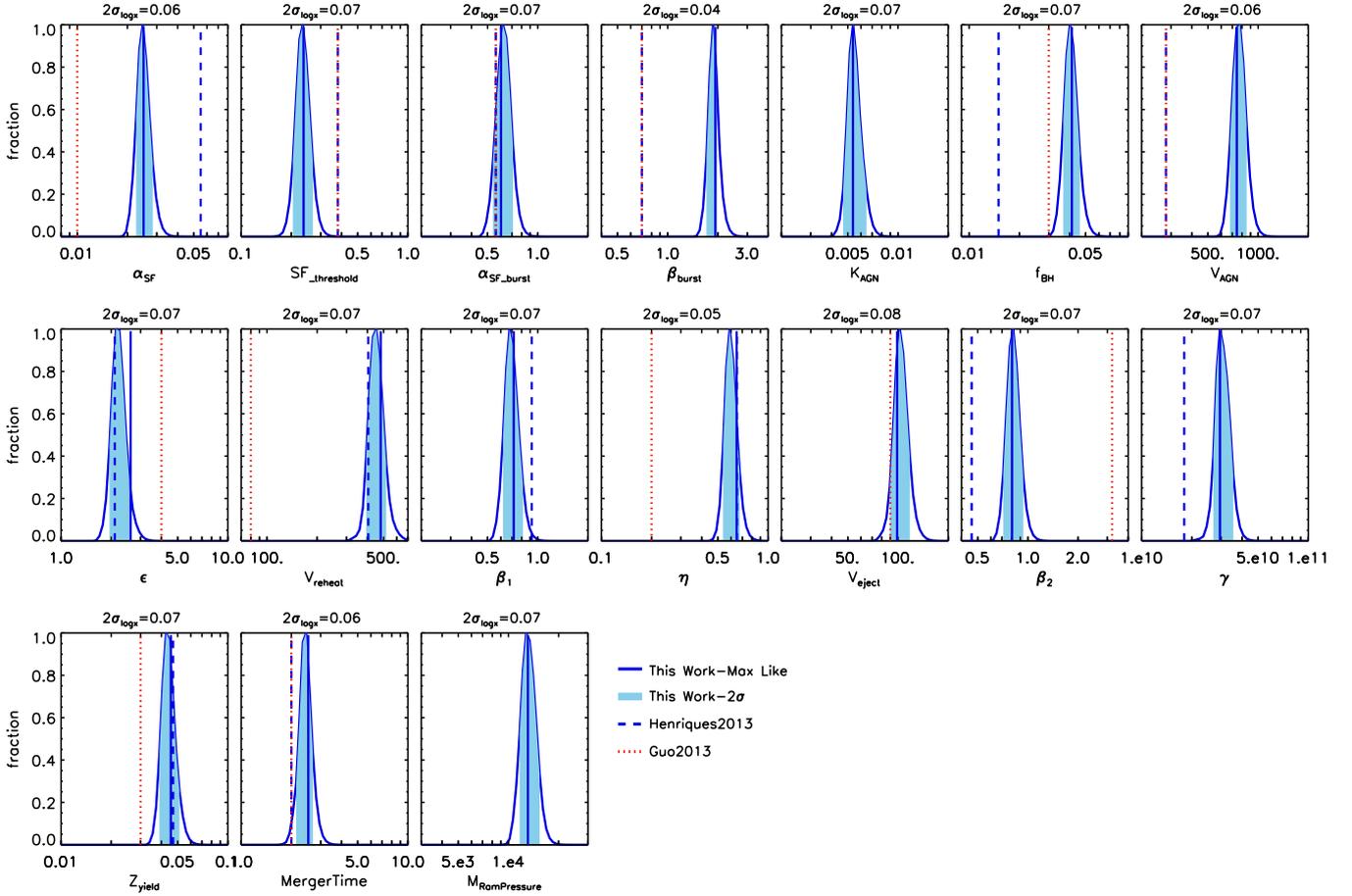

**Figure S9.** Shaded blue regions show the 1D, normalised posterior distributions of our model parameters when the model is constrained by observations of the abundance and passive fraction of galaxies as a function of stellar mass from $z = 3$ to 0. Straight lines represent values corresponding to our overall best-fitting model (solid blue lines) and to those of Henriques et al. (2013) (dashed blue lines) and Guo et al. (2013) (dotted red lines). The $x$-axis is plotted logarithmically in all cases.